\documentclass[twoside,journal]{IEEEtran}
    \usepackage{makecell}
    \usepackage{array}
    \usepackage{graphicx,amssymb,amsmath}
    \usepackage{multicol}
    \usepackage[noadjust]{cite}
    \usepackage{setspace}
    \usepackage{subfigure}
    \usepackage{graphicx}
    \usepackage{float}
    \usepackage {url}
    \usepackage{stfloats}
    \usepackage{amsthm,pifont}
    \usepackage{flushend}
    \usepackage{cases,subeqnarray}
    \usepackage{bm,multirow,bigstrut}
    \usepackage{amsmath, amsthm, amssymb}
    \usepackage{textcomp}
    \usepackage{latexsym,bm}
    \usepackage{booktabs}
    \usepackage{xcolor}
    \usepackage{mathtools}
    \usepackage{dsfont}
    \usepackage{extarrows}
    \usepackage{epsfig}
    \usepackage{epsfig}
    \usepackage{epstopdf}
    \usepackage[noend]{algpseudocode}
    \usepackage{hyperref}
    \usepackage{algorithmicx,algorithm}

    \theoremstyle{plain}

    \theoremstyle{plain}

    \usepackage{amsmath}

\begin{document}
    %----------------------------title&author&thanks----------------------------
    %\title{Contest Based Semantic Transmission of Sensing Information for Metaverse}
    % Contest Mechanism in Efficient Transmission of Sensing Information for Metaverse: A Semantic-aware Approach
    % Contest Mechanism in Semantic Transmission of Sensing Information to Build Metaverse
    %\title{Through the Wall Detection and Localization of Autonomous Mobile Device in 5G covered indoor environment}
    % \title{Towards Real-world Wireless Sensing: GAI assisted Human Flow Detection in Practical Communication Environments}
    \title{Generative Artificial Intelligence Assisted Wireless Sensing: Human Flow Detection in Practical Communication Environments}

    % Efficient Transmission of Sensing Information for Metaverse: Semantic-aware Approach and Contest Mechanism
    \author{Jiacheng Wang, Hongyang Du, Dusit~Niyato,~\IEEEmembership{Fellow,~IEEE}, Zehui Xiong, Jiawen Kang, Bo Ai,~\IEEEmembership{Fellow,~IEEE}, Zhu Han,~\IEEEmembership{Fellow,~IEEE}, and Dong In Kim,~\IEEEmembership{Fellow,~IEEE}

    \thanks{Jiacheng~Wang, Hongyang~Du, and Dusit Niyato are with the School of Computer Science and Engineering, Nanyang Technological University, Singapore 639798 (e-mail: jiacheng.wang@ntu.edu.sg, hongyang001@e.ntu.edu.sg, dniyato@ntu.edu.sg).}
    % \thanks{Mu Zhou is with School of Communications and Information Engineering, Chongqing University of Posts and Telecommunications, Chongqing, China, (zhoumu@cqupt.edu.cn).}
    % Corresponding author.
    \thanks{Zehui Xiong is with the Pillar of Information Systems Technology and Design, Singapore University of Technology and Design, Singapore (e-mail: zehui\_xiong@sutd.edu.sg).}
    \thanks{Jiawen Kang is with the School of Automation, Guangdong University of Technology (GDUT), China, and Key Laboratory of Intelligent Information Processing and System Integration of IoT (GDUT), Ministry of Education, Guangzhou 510006, China (e-mail: kavinkang@gdut.edu.cn).}

    \thanks{Bo Ai is with the State Key Laboratory of Rail Traffic Control and Safety and the Beijing Engineering Research Center of High-Speed Railway Broadband Mobile Communications, Beijing Jiaotong University, Beijing 100044, China (e-mail: boai@bjtu.edu.cn).}

    \thanks{Zhu Han is with the Department of Electrical and Computer Engineering, University of Houston, Houston, TX 77004 USA (e-mail: hanzhu22@gmail.com)}
    \thanks{Dong In Kim is with the Department of Electrical and Computer Engineering, Sungkyunkwan University, Suwon 16419, South Korea (email:dikim@skku.ac.kr).}
    %\thanks{Sumei Sun is with the Institute for Infocomm Research, Agency for Science, Technology and Research, Singapore 138632 (e-mail: sunsm@i2r.a-star.edu.sg).}
    % \thanks{Abbas Jamalipour is with the School of Electrical and Information Engineering, The University of Sydney, Australia, NSW 2006 (e-mail: a.jamalipour@ieee.org).}
    % \vspace{-1cm}
    }

    \maketitle
    %----------------------------abstract----------------------------

    \begin{abstract}
    Groundbreaking applications such as ChatGPT have heightened research interest in generative artificial intelligence (GAI). Essentially, GAI excels not only in content generation but also signal processing, offering support for wireless sensing. Hence, we introduce a novel GAI-assisted human flow detection system (G-HFD). Rigorously, G-HFD first uses the channel state information (CSI) to estimate the velocity and acceleration of propagation path length change of the human induced reflection (HIR). Then, given the strong inference ability of the diffusion model, we propose a unified weighted conditional diffusion model (UW-CDM) to denoise the estimation results, enabling detection of the number of targets. Next, we use the CSI obtained by a uniform linear array with wavelength spacing to estimate the HIR's time of flight and direction of arrival (DoA). In this process, UW-CDM solves the problem of ambiguous DoA spectrum, ensuring accurate DoA estimation. Finally, through clustering, G-HFD determines the number of subflows and the number of targets in each subflow, i.e., the subflow size. The evaluation based on practical downlink communication signals shows G-HFD's accuracy of subflow size detection can reach 91\%. This validates its effectiveness and underscores the significant potential of GAI in the context of wireless sensing.
    \end{abstract}
    %----------------------------keywords----------------------------
    %\newpage
    \vspace{-0.3cm}
    \IEEEpeerreviewmaketitle
    %---ction----------------------------
    \section{Introduction}
   %
    % The flow detection aims to acquire key information such as the number, location, and movement direction of human targets within a specified area, thereby facilitating crowd monitoring \cite{yu2022frequency}. Such a technology plays a crucial role in various scenarios such as shopping malls, tourist centers, and transportation hubs, as shown in Fig.~\ref{application}. For instance, in shopping mall, sellers can utilize the flow detection system to gather information about the number of customers and flow patterns at different times and locations, allowing for product inventory optimization and staff resource allocation~\cite{huang2019pedestrian}. In transportation hubs like subway stations, the flow detection aids in congestion control and management~\cite{duran2021crowd}. In emergency situations, it also helps administrators ensure the safety levels of individuals in various areas and make appropriate emergency responses.

\begin{figure*}[htp]
    \centering
    \includegraphics[width=1\textwidth]{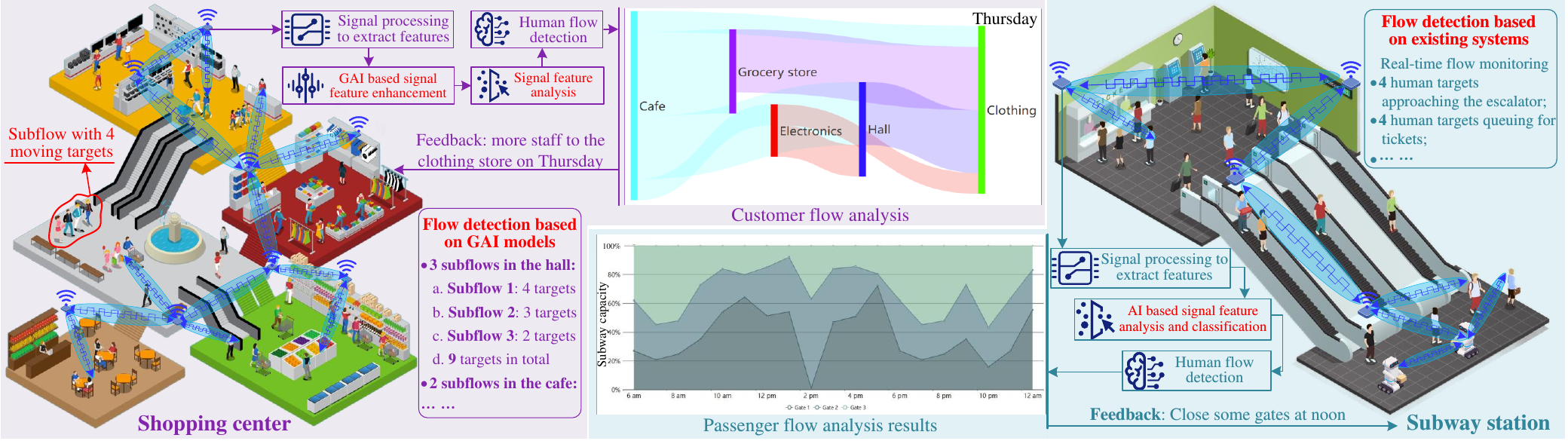}
    \caption{The application of flow detection and the differences between the proposed method and existing approaches. For instance, in subway stations, such detection systems can continuously monitor the number of passengers and their flow across different areas, facilitating more effective crowd congestion control and management. Existing systems use conventional AI models to analyze the extracted signal features for flow detection, with performance being limited by the quality of signal features. Different from them, the proposed method uses GAI to enhance the signal features, thereby supporting fine-grained human flow detection.}
    \label{application}
    \vspace{-0.5cm}
    \end{figure*}

    The ubiquitous penetration and inter-connectivity of trillions of smart wireless devices are ushering us into the era of the Artificial Intelligence of Things (AIoT)~\cite{zhang2020empowering}. Unlike before, in this era, the advancement in AI and wireless technologies is revolutionizing wireless networks from mere communication mediums into pervasive sensing platforms~\cite{bai2019camera}. Such a paradigm shift paves the way for numerous contactless applications~\cite{zhu2023positioning,wang2021leveraging,sun2024joint,wang2023semantic}, among which human flow detection stands as a critical area of interest. Human flow detection technology seeks to gather essential information, including the number, location, and movement of people within a specified area, thereby facilitating crowd monitoring \cite{yu2022frequency}. Such a technology plays a crucial role in various scenarios such as shopping malls and transportation hubs, as shown in Fig.~\ref{application}.

    Traditional signal processing-based flow detection approaches are facing performance bottlenecks in distinguishing human-relevant features from entangled irrelevant features within the signal~\cite{yang2023slnet}. Therefore, AI technologies are introduced to enhance the detection performance~\cite{zhou2020wiflowcount, hao2023toward,liu2022sensor}. \textcolor{black}{For instance, in~\cite{zhou2020wiflowcount}, the authors constructed the spectrogram of Doppler shifts and then used a rotation and segmentation algorithm to divide the spectrogram into the subspectrograms of its subflows. On this basis, the number of people in each subflow is estimated via the convolutional neural network (CNN). In another work~\cite{hao2023toward}, the authors defined the sequential spatial-temporal matrix and input it into a recurrent neural network (RNN) to mine the spatial-temporal correlations among crowd features, thereby achieving an estimation of target counts. The evaluation shows it reduced the counting error rate from 22.54\% to 13.44\% compared with several state-of-the-art methods. Besides, authors in~\cite{liu2022sensor} used the wavelet transform based method to denoise the CSI and then extract four features that can depict the relationship between the number of people and data fluctuation. After that, they trained a one-vs-rest support vector machine (SVM) model to realize the crowd counting with the accuracy of 87.2\%.}

    In these systems~\cite{mizutani2020towards, sharma2022optimised}, AI technologies focus on signal feature analysis and classification, yet they fail to improve the quality of signal feature parameters. Therefore, they still face certain limitations. For instance, the array designs of the commercial devices often do not align with the assumptions made in existing research, particularly regarding antenna spacing~\cite{wang2023through,qian2018widar2,li2017indotrack,xie2019md,liu2021hiloc}, which complicates the estimation of parameters such as the direction of arrival (DoA). These constraints make the obtained signal features insufficient for more fine-grained human flow detection, such as identifying the number and size of subflows. The key to overcome these limitations and advancing practical human flow detection lies in enhancing the quality of signal features, for which generative AI (GAI) presents a promising solution~\cite{du2023ai}.

     Unlike traditional AI technologies for signal feature analysis and classification, GAI has stronger data processing and inferencing capabilities~\cite{sun2020generative, saidutta2021joint}. It can capture latent relationships between complex distributions across various dimensions and spaces, and then leverage relationships to generate data that fulfills specific requirements based on given conditions~\cite{guo2022systematic}. \textcolor{black}{For example, the diffusion model~\cite{yang2022diffusion} is one of the representative GAI models. It starts with adding Gaussian noise to disrupt the training samples and then learns to remove this noise in the reverse process to generate new samples~\cite{van2023generative}. This working principle makes it particularly suitable for applications such as image denoising~\cite{kulikov2023sinddm}, resolution enhancement~\cite{li2022srdiff}, specified content generation~\cite{wang2024unified}, and so forth, where it has demonstrated impressive performance. In this context, the given image is a dataset with a particular distribution, which is essentially the same as the signal features, such as frequency spectrum, used for human flow detection~\cite{di2016trained}. Therefore, we can leverage the powerful generative capabilities of diffusion model to denoise, repair, and enhance the extracted signal features, under specific conditions. This goes beyond the analysis and classification functions of traditional AI technologies in current human flow detection systems, paving the way for the practical deployment of human flow detection.}

    Therefore, in this paper, we introduce a GAI assisted human flow detection system (G-HFD). Specifically, G-HFD first utilizes the channel state information (CSI) measurements in time domain to estimate the velocity and acceleration of propagation path length change (PPLC) of the human induced reflection (HIR). Then, we propose a unified weighted conditional diffusion model (UW-CDM) to denoise the estimation results, facilitating the detection of the number of targets in human flow. After that, the CSI obtained by a uniform linear array (ULA) with the antenna spacing equal to a wavelength is used to estimate the DoA and ToF of the HIR. During this process, UW-CDM is used to solve the problem of ambiguous DoA spectrum, thereby ensuring the effective DoA estimation. Finally, the obtained signal parameters are clustered to determine the number of subflows and the subflow size, i.e., the number of targets in each subflow. Our evaluation of G-HFD, using downlink signals in practical communication scenarios, confirms its effectiveness. The contributions of this paper are summarized as follows.

    \begin{itemize}
    \item We propose UW-CDM, which allows us to denoise the spectrum obtained by velocity and acceleration estimation. This enables the identification of the total number of human targets and each one's velocity and acceleration. Such a method utilizes CSI in the time domain, and hence the resolution is not limited by signal bandwidth or the number of antennas.

    \item We employ the UW-CDM to generate the DoA spectrum based on the CSI measurements obtained by the ULA, where antenna spacing is one wavelength. This enables the acquisition of HIR's DoA when the antenna spacing is greater than half a wavelength, strongly supporting the practical application of the proposed G-HFD.

    \item We analyze the obtained DoA, ToF, and velocity of the HIR by employing clustering techniques to achieve human flow detection. This detection includes the number of subflows and the subflow size (i.e., the number of targets in each subflow), which are hard to achieve for existing works.

    \item We conduct comprehensive tests in practical communication scenarios by using downlink signals. Experimental results demonstrate that, when the user equipment (UE) downloads files, the G-HFD's accuracy of subflow size detection can reach 91\%, which not only validates its effectiveness but also illustrates the potential of GAI in enhancing wireless sensing capabilities.
    \end{itemize}

    This paper is organized as follows. Section II is the system design, including the system overview, velocity and acceleration estimation, and so forth. Section III presents the implementation and evaluation, and Section IV is the conclusion.
    \begin{figure*}[htp]
    \centering
    \includegraphics[width=1\textwidth]{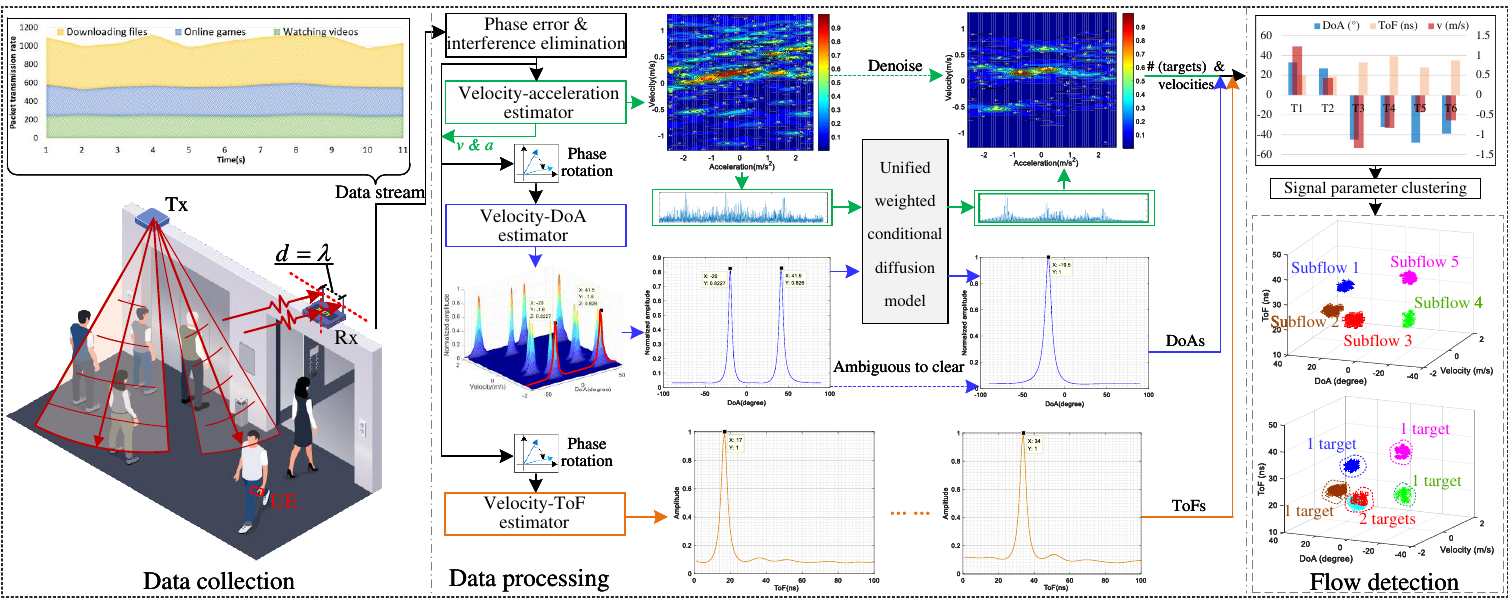}
    \caption{The framework of the proposed G-HFD. Here, the Tx is the signal transmitter, Rx is the receiver, and UE is the mobile phone. In a practical communication scenario, the Rx first employs a ULA with the antenna spacing equal to one wavelength to capture the downlink signals and extract CSI. Then, G-HFD utilizes the obtained CSI to estimate the signal parameters of the HIR, including acceleration, velocity, DoA, and ToF. In this process, UW-CDM is used to denoise the velocity and acceleration estimation results and generate the clear DoA spectrum, so as to accurately estimate the signal parameters. Finally, through clustering, G-HFD realizes the human flow detection, which includes the number of human targets and subflows, and the subflow size.}
    \label{SYST}
    \vspace{-0.5cm}
    \end{figure*}

    \section{System Design}
    \subsection{System Overview}
    As shown in Fig.~\ref{SYST}, the key objective of G-HFD is to detect the flow via effective estimation and analysis of signal parameters across various domains. First, leveraging the CSI data stream in the time domain, G-HFD estimates the velocity and acceleration of PPLC of the reflections induced by the moving human targets in the monitored area. This estimation yields a two-dimensional spectrum of velocity and acceleration (denoted as V-A spectrum). Subsequently, the proposed UW-CDM is trained to denoise the V-A spectrum, thereby facilitating an estimation of the number of moving human targets. Following this, G-HFD estimates the DoA and ToF for each HIR based on the CSI collected by a ULA with the antenna spacing equal to one wavelength. In this process, UW-CDM is employed again to address the issue of ambiguous DoA spectrum caused by antenna spacing greater than half a wavelength. Finally, the clustering is conducted on the obtained velocity, DoA, and ToF, thereby identifying the number of subflows and the subflow size.
    % \vspace{-0.5cm}
    \subsection{Velocity and Acceleration Estimation}
    Consider a real-world communication scenario shown in Fig.~\ref{SYST}. Here, a user's mobile phone connects to an access point (AP), denoted as Tx, to access the Internet, forming a communication link. At the same time, another AP, denoted as Rx, is deployed to receive and process the downlink signal for the human flow detection. The Rx is equipped with $M+1$ antennas, with one of them designated for receiving the direct signals from Tx to Rx, which are the reference signals. The remaining $M$ antennas form a ULA, with an antenna spacing of one wavelength, to receive signals in free space, which are treated as surveillance signals. Therefore, the CSI of surveillance signal corresponding to the $k$-th subcarrier and $m$-th antenna of the ULA can be denoted as
    \begin{align}\label{eq1}
    {h_{m,k}} = \sum\limits_{b = 1}^B {\alpha _{m,k}^{\left[ b \right]}{e^{ - j2\pi {f_k}\tau _m^{\left[ b \right]}}}{e^{ - j2\pi \varepsilon }}}  + {n_{m,k}},
    \end{align}
    where $B$ is the total number of propagation signals, including the direct signals (from Tx to Rx) and reflection signals, $m = 1,{\rm{ }} \ldots {{, }\ }M$, $\alpha _{m,k}^{\left[ b \right]}$ is the attenuation, ${f_k}$ is the frequency of the $k$-th subcarrier, $\tau _m^{\left[ b \right]}$ is the signal ToF, ${e^{ - j2\pi \varepsilon }}$ is the phase error introduced by the timing jitter during synchronization between Tx and Rx, and ${n_{m,k}}$ is the noise. Similarly, the CSI of reference signal corresponding to the $k$-th subcarrier is denoted as
    \begin{align}\label{eq2}
    {h_{re,k}} = \sum\limits_{b' = 1}^{B'} {\alpha _{re,k}^{\left[ {b'} \right]}{e^{ - j2\pi {f_k}\tau _{re}^{\left[ {b'} \right]}}}{e^{ - j2\pi \varepsilon }}}  + {n_{re,k}},
    \end{align}
    where $B'$ is the total number of propagation signals, $\alpha _{re,k}^{\left[ {b'} \right]}$ is the attenuation, $\tau _{re}^{\left[ {b'} \right]}$ is the signal ToF, and ${n_{{re},k}}$ is the noise. To accurately estimate the velocity and acceleration of the PPLC corresponding to the each HIR, the complex conjugate multiplication is first conducted to remove the phase error
    \begin{align}\label{eq3}
    {h_{m\bar {e},k}} = {h_{m,k}} {{\bar h}_{re,k}} {\rm{ = }}\sum\limits_{r = 1}^R {\alpha _{m\bar e,k}^{\left[ r \right]}{e^{ - j2\pi {f_k}\tau _{m\bar {e}}^{\left[ r \right]}}}}  + \Gamma  + \Gamma ' + {n_{m\bar {e},k}},
    \end{align}
    where ${{\bar h}_{re,k}}$ represents the conjugate of ${h_{re,k}}$, $R = BB'$, $\alpha _{m\bar {e},k}^{\left[ r \right]} = \alpha _{m,k}^{\left[ b \right]} \times \alpha _{re,k}^{\left[ {b'} \right]}$, $\tau _{m\bar {e}}^{\left[ r \right]}$ is the ToF difference, ${n_{m\bar e,k}}$ is the product of two noise terms, $\Gamma $ and $\Gamma'$ are the results of cross-multiplication of signal and noise. In ${h_{m\bar e,k}}$, the product of the direct signals from ${h_{m,k}}$ and ${h_{re,k}}$, denoted as the direct component, has the strongest amplitude. Then, the product of the direct signals and the reflection signals, denoted as a cross component, has a lower amplitude than that of the direct component. Besides, ${h_{m\bar e,k}}$ includes the product of the reflection signals from ${h_{m,k}}$, and ${h_{re,k}}$ (denoted as the reflection component), and the product of noise and the signal (denoted as the noise component). These two components have smaller amplitudes than those of the direct and cross components. As can be seen, except for the noise component, the signals contained in the other components no longer contain phase errors, laying the foundation for velocity and acceleration estimation. Hence, let ${h_{m\bar e,k}}$ be the observation at time 0 and the $r$-th signal is the HIR with the velocity and acceleration of ${v_r}$ and ${a_r}$, respectively, then at time $\Delta t$, the PPLC can be calculated as
    \begin{align}\label{eq4}
    d\left( {\Delta t} \right) = {v_r}\Delta t + \frac{1}{2}{a_r}{\left( {\Delta t} \right)^2},
    \end{align}
    which causes a phase shift of
    \begin{align}\label{eq5}
    P{h_{k,r}} = {e^{ - j2\pi {f_k}\frac{{d\left( {\Delta t} \right)}}{c}}} = {e^{ - j2\pi {f_k}\left( {\frac{{{v_r}\Delta t}}{c} + \frac{{{a_r}{{\left( {\Delta t} \right)}^2}}}{{2c}}} \right)}}.
    \end{align}
    Therefore, at time $\Delta t$, the signal contained in the cross component\footnote{The signals in direct component and reflection component do not include terms related to ${\Delta t}$, therefore, we do not discuss them here.} in ${h_{m\bar e,k}}$ can be expressed as
    \begin{align}\label{eq6}
    {h'_{m\bar e,k}}\left( {\Delta t} \right) = \sum\limits_{r' = 1}^{R'} {\alpha _{m\bar e,k}^{\left[ {r'} \right]}{e^{ - j2\pi {f_k}\left[ {\tau _{m\bar e}^{\left[ {r'} \right]} + \frac{{{v_{r'}}\Delta t}}{c} + \frac{{{a_{r'}}{{\left( {\Delta t} \right)}^2}}}{{2c}}} \right]}}},
    \end{align}
    where $R'$ represents the total number of signals contained in the cross component.

    The analysis above reveals that the phase of the reflection signal introduced by the moving human target is a function of the PPLC velocity and acceleration, as well as ${\Delta t}$. This relationship allows us to estimate the velocity and acceleration based on the CSI data stream in time domain. However, as previously mentioned, in ${h_{m\bar e,k}}$, the direct component, with the velocity and acceleration of zero, has the strongest amplitude. This causes the cross component to be easily submerged in the direct component, causing interference to the estimation. Moreover, the velocity and acceleration of the reflection component are also zero, exacerbating such effect. Therefore, it is necessary to eliminate the interference before estimation. Given that the phase of the signals in direct and reflection components does not change over time, implying a frequency of zero, we first obtain $W$ observations from the CSI stream in the time domain to construct the following matrix
    \begin{align}\label{eq7}
    {{\bf{H}}_{m\bar e,k}} = \left[ {{h_{m\bar e,k}}\left( 0 \right),{h_{m\bar e,k}}\left( {\Delta t} \right), \ldots ,{h_{m\bar e,k}}\left( {\left( {W - 1} \right)\Delta t} \right)} \right].
    \end{align}
    Through the Fast Fourier Transform (FFT), the ${{\bf{H}}_{m\bar e,k}}$ is converted into the frequency domain, yielding ${{\bf{H'}}_{m\bar e,k}}$. After that, we nullify the values at the frequency of 0 in ${{\bf{H'}}_{m\bar e,k}}$ to eliminate interference. Subsequently, the Inverse Fast Fourier Transform (IFFT) is employed to revert it to the time domain for the estimation.

    Let ${{\bf{S}}_{m,k}} = [{s_{m,k}}\left( 0 \right),{\rm{ }}{s_{m,k}}\left( {\Delta t} \right),{\rm{ }} \ldots {\rm{, }}{s_{m,k}}\left( {\left( {W - 1} \right)\Delta t} \right)]$ be the data stream transformed back to the time domain, which contains cross component, $\Gamma $, $\Gamma' $, and noise. Among these, the cross component exhibits the largest amplitude and is free from phase errors. Hence, we simplify $\alpha _{m\bar e,k}^{\left[ {r'} \right]}$ to $\alpha _k^{\left[ {r'} \right]}$ and derive the estimation of velocity and acceleration based on the cross component. More concretely, we first calculate the parametric symmetric instantaneous auto-correlation function of ${{\bf{S}}_{m,k}}$ as
    \begin{align}\label{eq8}
    &A_{{s_{m,k}}}^C\left( {\Delta t,\tau } \right) = {s_{m,k}}\left( {\Delta t + \frac{{\tau  + {t_d}}}{2}} \right){{\bar s}_{m,k}}\left( {\Delta t - \frac{{\tau  + {t_d}}}{2}} \right) \notag \\
    &= \underbrace {\sum\limits_{r' = 1}^{R'} {{{\left( {\alpha _k^{\left[ {r'} \right]}} \right)}^2}{e^{ - j2\pi \left( {\tau  + {t_d}} \right)\frac{{{f_k}}}{c}\left( {{v_{r'}} + {a_{r'}}}\Delta t \right)}}} }_{auto{\ }term}\notag \\
    &+ \underbrace {\sum\limits_{r' = 1}^{R' - 1} {\sum\limits_{r'' = r' + 1}^{R'} {\left[ {A_{s_{m,k}^{\left[ {r'} \right]}s_{m,k}^{\left[ {r''} \right]}}^C\left( {\Delta t,\tau } \right) + A_{s_{m,k}^{\left[ {r''} \right]}s_{m,k}^{\left[ {r'} \right]}}^C\left( {\Delta t,\tau } \right)} \right]} } }_{cross{\ }term},
    \end{align}
    where the second row is the auto term, the third row is the cross term, and ${t_d}$ is the constant time-delay corresponds to a scaling operator. One can see from the $A_{{s_{m,k}}}^C\left( {\Delta t,\tau } \right)$ that the time variable $\Delta t$ and lag variable $\tau$ are coupled in the terms of the exponential phase. Therefore, the keystone transformation~\cite{lv2011lv} is used to re-scale the time axis for each lag. Specifically, the transformation is defined as
    \begin{align}\label{eq9}
    \Psi \left[ {A_{{s_{m,k}}}^C\left( {\Delta t,\tau } \right)} \right] = A_{{s_{m,k}}}^C\left( {\frac{g}{{z\left( {\tau  + {t_d}} \right)}},\tau } \right).
    \end{align}

    Leveraging this definition, we perform the transformation on the auto term and cross term in~(\ref{eq8}), yielding~(\ref{eq10}) and (\ref{eq11}), shown in the bottom of next page, respectively, where ${\mathop{\rm Re}\nolimits} \left(  \cdot  \right)$ indicates the operation of taking the real part.
    \newcounter{mycount}
    \begin{figure*}[b]
	\normalsize
	\setcounter{mycount}{\value{equation}}
	\hrulefill
	\vspace*{4pt}
\begin{align}\label{eq10}
    \Psi &\left[ {\sum\limits_{r' = 1}^{R'} {{{\left( {\alpha _k^{\left[ {r'} \right]}} \right)}^2}{e^{ - j2\pi \left( {\tau  + {t_d}} \right)\frac{{{f_k}}}{c}\left( {{v_{r'}} + {a_{r'}}}\Delta t \right)}}} } \right]
    = \sum\limits_{r' = 1}^{R'} {{{\left( {\alpha _k^{\left[ {r'} \right]}} \right)}^2}{e^{ - j2\pi \frac{{{f_k}}}{c}\left[ {{v_{r'}}\left( {\tau  + {t_d}} \right) + \frac{{{a_{r'}}}}{z}g} \right]}}}
    \end{align}
\end{figure*}
% \addtocounter{equation}{1}	

    \begin{figure*}[b]
	\normalsize
	\setcounter{mycount}{\value{equation}}
	\hrulefill
	\vspace*{4pt}
    \begin{align}\label{eq11}
    &\Psi \left[ {\sum\limits_{r' = 1}^{R' - 1} {\sum\limits_{r'' = r' + 1}^{R'} {\left[ {A_{s_{m,k}^{\left[ {r'} \right]}s_{m,k}^{\left[ {r''} \right]}}^C\left( {\Delta t,\tau } \right) + A_{s_{m,k}^{\left[ {r''} \right]}s_{m,k}^{\left[ {r'} \right]}}^C\left( {\Delta t,\tau } \right)} \right]} } } \right]\notag \\
   & = 2\alpha _k^{\left[ {r'} \right]}\alpha _k^{\left[ {r''} \right]}{e^{ - j2\pi \frac{{{f_k}}}{c}\left[ {\left( {{v_{r'}}{\rm{ + }}{v_{r''}}} \right)\left( {\tau  + {t_d}} \right) - \left( {{a_{r'}}{\rm{ + }}{a_{r''}}} \right)\frac{g}{z}} \right]}}
    {\mathop{\rm Re}\nolimits} \left[ \begin{array}{l}
    {e^{ - j2\pi \frac{{{f_k}}}{c}\frac{{2g\left( {{v_{r'}} - {v_{r''}}} \right)}}{{z\left( {\tau  + {t_d}} \right)}}}}
    \times {e^{ - j2\pi \frac{{{f_k}}}{c}\left[ {\frac{{\left( {{a_{r'}}{\rm{ + }}{a_{r''}}} \right){t^2}}}{{{z^2}{{\left( {\tau  + {t_d}} \right)}^2}}} + \frac{{\left( {{a_{r'}}{\rm{ - }}{a_{r''}}} \right){{\left( {\tau  + {t_d}} \right)}^2}}}{4}} \right]}}
    \end{array} \right]
    \end{align}
    \end{figure*}
    % \addtocounter{equation}{1}	
    After de-coupling, the two-dimensional FFT transformation is applied to $\Psi \left[ {A_{{s_{m,k}}}^C\left( {\Delta t,\tau } \right)} \right]$ with respect to $g$ and $\tau$, obtaining
    \begin{align}\label{eq12}
    {G_{{s_{m,k}}}}\left( {v,a} \right) &= {{{\cal F}}_\tau }\left( {{{{\cal F}}_g}\left( {\Psi \left[ {A_{{s_{m,k}}}^C\left( {\Delta t,\tau } \right)} \right]} \right)} \right) \notag \\
    &= \underbrace {\sum\limits_{r' = 1}^{R'} {{G_{s_{m,k}^{\left[ {r'} \right]}}}\left( {v,a} \right)} }_{auto \ term} \notag \\
    &+ \underbrace {\sum\limits_{r' = 1}^{R' - 1} {\sum\limits_{r'' = r' + 1}^{R'} {{G_{s_{m,k}^{\left[ {r'} \right]}s_{m,k}^{\left[ {r''} \right]}}}\left( {v,a} \right)} } }_{cross \ term},
    \end{align}
    where the second and third rows present the transformation result of the auto term and cross term, respectively. Specifically, let $\tau ' = \tau  + {t_d}$ and conduct Fourier transform of the $r'$-th auto term in (\ref{eq10}) with respect to $g$, we have
    \begin{align}\label{eq13}
    {G_{{s_{m,k}}}}\left( {\tau ,a} \right) &= {{\cal F}_g}\left[ {{{\left( {\alpha _k^{\left[ {r'} \right]}} \right)}^2}{e^{ - j2\pi \frac{{{f_k}}}{c}\left[ {{v_{r'}}\left( {\tau  + {t_d}} \right) + \frac{{{a_{r'}}}}{z}g} \right]}}} \right]\notag \\
    &= {\left( {\alpha _k^{\left[ {r'} \right]}} \right)^2}{e^{ - j2\pi \frac{{{f_k}}}{c}{v_{r'}}\tau '}} \notag \\
    &\times \int_{ - \infty }^\infty  {{e^{ - j2\pi \frac{{{f_k}}}{c}\left( {a + \frac{{{a_{r'}}}}{z}} \right)g}}} dg\notag \\
    &= {\left( {\alpha _k^{\left[ {r'} \right]}} \right)^2}{e^{ - j2\pi \frac{{{f_k}}}{c}{v_{r'}}\tau '}}\delta \left( {a + \frac{{{a_{r'}}}}{z}} \right),
    \end{align}
    where $\delta \left(  \cdot  \right)$ is the Dirac delta function. On this basis, performing the Fourier transform on (\ref{eq13}) with respect to $\tau$ leads to
    \begin{align}\label{eq14}
    {G_{{s_{m,k}}}}\left( {v,a} \right) &= {{\cal F}_\tau }\left[ {{{\left( {\alpha _k^{\left[ {r'} \right]}} \right)}^2}{e^{ - j2\pi \frac{{{f_k}}}{c}{v_{r'}}\tau '}}\delta \left( {a + \frac{{{a_{r'}}}}{z}} \right)} \right]\notag \\
    &= {\left( {\alpha _k^{\left[ {r'} \right]}} \right)^2}\delta \left( {v + {v_{r'}}} \right)\delta \left( {a + \frac{{{a_{r'}}}}{z}} \right) \notag \\
    &\times \exp \left( { - j2\pi \frac{{{f_k}}}{c}{t_d}{v_{r'}}} \right).
    \end{align}

\begin{figure*}[htp]
    \centering
    \includegraphics[width=1\textwidth]{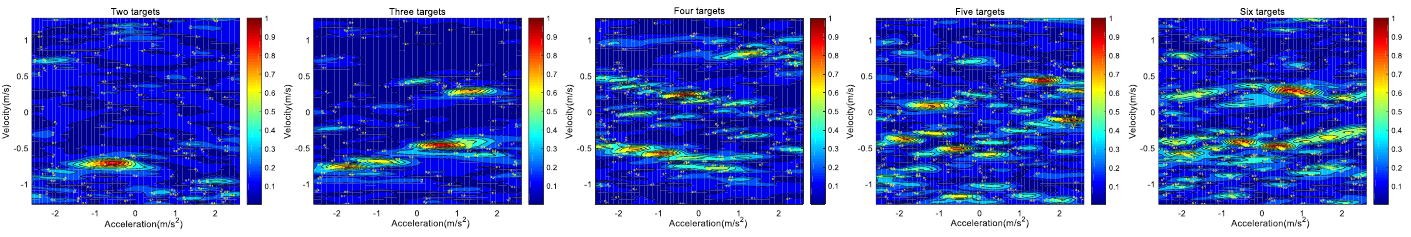}
    \caption{The V-A spectra extracted via the velocity and acceleration estimation when different numbers of moving human targets appear in the monitored area.}
    \label{v-a}
    % \vspace{-0.5cm}
    \end{figure*}

    In (\ref{eq14}), each auto term in (\ref{eq10}) can be modeled as a Dirac delta function, resulting in an impulse on the generated V-A spectrum, and the velocity and acceleration can be obtained from the coordinates of the impulse. Figure \ref{v-a} presents an estimation result based on the CSI collected in a real-world scenario when different numbers of moving targets appear in the monitored area. As can be seen, for instance, when two moving targets appear, two distinct peaks can be observed from the V-A spectrum. Through these peaks, we can identify the number of moving human targets and obtain their corresponding velocities and accelerations. However, the Fourier transform results of the cross terms cause interference in the V-A spectrum. As the number of moving targets increases, the interference makes it challenging to identify peaks corresponding to moving human targets in the V-A spectrum. Therefore, we introduce UW-CDM to reduce the noise in the V-A spectrum, facilitating the estimation of velocity, acceleration, and the number of the moving targets.
    \subsection{Unified Weighted Conditional Diffusion Model}
    % \vspace{-0.5cm}
    The diffusion model includes the forward process and reverse process ~\cite{du2023ai}. During the forward process, it adds noise to disturb the training data. In the reverse process, it predicts the noise to train the denoising network~\cite{luo2022understanding}. Once completed, it can generate new data via the denoising network. The generative diffusion model has demonstrated powerful capabilities in image denoising~\cite{kulikov2023sinddm}, restoration~\cite{zhu2023denoising}, synthesis~\cite{dhariwal2021diffusion}, and even channel estimation~\cite{arvinte2022mimo}. Inspired by this, we propose UW-CDM. The core of UW-CDM is to constrain the diffusion model by adding conditions, forcing it to generate desired data, such as the V-A spectrum with less noise, to support human flow detection.

    In the diffusion model, the forward process is defined as a Markov chain. Let ${{\bf{x}}_0}$ be the original data, in the forward diffusion process, $T$ rounds of the Gaussian noise are added to ${{\bf{x}}_0}$, yielding ${{\bf{x}}_t}$, which can be represented as
    \begin{align}\label{eq15}
    q\left( {{{\bf{x}}_{1:T}}|{{\bf{x}}_0}} \right) &= \prod\limits_{t = 1}^T {q\left( {{{\bf{x}}_t}|{{\bf{x}}_{t - 1}}} \right)} \notag \\
    &= \prod\limits_{t = 1}^T {{{\cal N}}\left( {{{\bf{x}}_t};\sqrt {1 - {\beta _t}} {{\bf{x}}_{t - 1}},{\beta _t}{\bf{I}}} \right)},
    \end{align}
    where
    \begin{equation}\label{eq16}
    q\left( {{{\bf{x}}_t}|{{\bf{x}}_{t - 1}}} \right) = {{\cal N}}\left( {{{\bf{x}}_t};{{\bm{\mu}}_t}= \sqrt {1 - {\beta _t}} {{\bf{x}}_{t - 1}},{\beta _t}{\bf{I}}} \right),
    \end{equation}
    ${\beta _t}$ is the variance, and ${\bf{I}}$ is the identity matrix. Building on this, we first add the condition ${\bf{u}}$ to the forward process. Given that the diffusion model is inherently a Markov chain, the value at any step $t$ depends solely on its previous state. This implies that the forward diffusion conditional probability at any step $t$ is independent of ${\bf{u}}$. Hence, we have
    \begin{equation}\label{eq17}
    q'\left( {{{\bf{x}}_t}|{{\bf{x}}_{t - 1}},{\bf{u}}} \right) = q\left( {{{\bf{x}}_t}|{{\bf{x}}_{t - 1}}} \right).
    \end{equation}
    On this basis, we can further obtain that
    \begin{align}\label{eq18}
    q'\left( {{{\bf{x}}_t}|{{\bf{x}}_{t - 1}}} \right)& = \int_u {q'\left( {{{\bf{x}}_t},{\bf{u}}|{{\bf{x}}_{t - 1}}} \right)d} {\bf{u}} \notag \\
    &= \int_u {q'\left( {{{\bf{x}}_t}|{\bf{u}},{{\bf{x}}_{t - 1}}} \right)q'\left( {{\bf{u}}|{{\bf{x}}_{t - 1}}} \right)d} {\bf{u}} \notag \\
    &= \int_u {q\left( {{{\bf{x}}_t}|{{\bf{x}}_{t - 1}}} \right)q'\left( {{\bf{u}}|{{\bf{x}}_{t - 1}}} \right)d} {\bf{u}}\notag \\
    &= q\left( {{{\bf{x}}_t}|{{\bf{x}}_{t - 1}}} \right) = q'\left( {{{\bf{x}}_t}|{{\bf{x}}_{t - 1}},{\bf{u}}} \right).
    \end{align}
    Following a similar way, the joint distribution can be obtained
    \begin{align}\label{eq19}
    q'\left( {{{\bf{x}}_{1:T}}|{{\bf{x}}_0}} \right) &= \int_u {q'\left( {{{\bf{x}}_{1:T}},{\bf{u}}|{{\bf{x}}_0}} \right)d} {\bf{u}} \notag \\
    &= \int_u {q'\left( {{\bf{u}}|{{\bf{x}}_0}} \right)q'\left( {{{\bf{x}}_{1:T}}|{{\bf{x}}_0},{\bf{u}}} \right)d} {\bf{u}}\notag \\
    &= \int_u {q'\left( {{\bf{u}}|{{\bf{x}}_0}} \right)\prod\limits_{t = 1}^T {q'\left( {{{\bf{x}}_t}|{{\bf{x}}_{t - 1}},{\bf{u}}} \right)} d} {\bf{u}}\notag \\
    &= \prod\limits_{t = 1}^T {q'\left( {{{\bf{x}}_t}|{{\bf{x}}_{t - 1}}} \right)} = q\left( {{{\bf{x}}_{1:T}}|{{\bf{x}}_0}} \right).
    \end{align}

    From (\ref{eq15}) to (\ref{eq19}), the derivation process proves that the condition ${\bf{u}}$ has no effect on the forward diffusion process. However, it is completely different with the reverse diffusion process. In the reverse process, the diffusion model starts with Gaussian noise and generates samples through $T$ denoising steps. If the reverse distribution $q\left( {{{\bf{x}}_{t - 1}}\left| {{{\bf{x}}_t}} \right.} \right)$ can be obtained, then we can effectively execute the reverse diffusion process and obtain samples from $q\left( {{{\bf{x}}_0}} \right)$. However, obtaining $q\left( {{{\bf{x}}_{t - 1}}\left| {{{\bf{x}}_t}} \right.} \right)$ involves computing the data distribution, which is, in practice, intractable. Thereby, in the reverse process, $q\left( {{{\bf{x}}_{t - 1}}\left| {{{\bf{x}}_t}} \right.} \right)$ is estimated using the parametric model
    \begin{align}\label{eq20}
    {p_\theta }\left( {{{\bf{x}}_{t - 1}}|{{\bf{x}}_t}} \right) = {{\cal N}}\left( {{{\bf{x}}_{t - 1}};{{\bm{\mu }}_\theta }\left( {{{\bf{x}}_t},t} \right),{{\bm{\Sigma }}_\theta }\left( {{{\bf{x}}_t},t} \right)} \right).
    \end{align}
    Following this, the trajectory from \(x_T\) to \(x_0\) is denoted as
    \begin{align}\label{eq21}
    {p_\theta }\left( {{{\bf{x}}_{0:T}}} \right) = {p_\theta }\left( {{{\bf{x}}_T}} \right)\prod\limits_{t = 1}^T {{p_\theta }\left( {{{\bf{x}}_{t - 1}}|{{\bf{x}}_t}} \right)},
    \end{align}
    and the loss function can be expressed as

    \begin{align}\label{eq22}
    L\left( \theta  \right){\rm{ = }}{\mathbb{E}_{{{\bf{x}}_0},{\bm{\varepsilon }} \sim N\left( {0,{\bf{I}}} \right),t}}\left[ {{{\left\| {{{\bm{\varepsilon }}_\theta }\left( \begin{array}{l}
    \sqrt {{{\bar \alpha }_t}} {{\bf{x}}_0}\\
    + \sqrt {1 - \sqrt {{{\bar \alpha }_t}} } {\bm{\varepsilon }},t
    \end{array} \right) - {\bm{\varepsilon }},t} \right\|}^2}} \right],
    \end{align}
    where ${\bar \alpha _t} = \prod\limits_{i = 0}^t {\left( {1 - {\beta _i}} \right)}$. Here, $L\left( \theta  \right)$ is a weighted form of the evidence lower bound (ELBO) $ - {L_\theta }\left( {{{\bf{x}}_0}} \right) \le \log {p_\theta }\left( {{{\bf{x}}_0}} \right)$, where $ {L_\theta }\left( {{{\bf{x}}_0}} \right)$ is shown in (23) at the bottom of this page and ${L_T}\left( {{{\bf{x}}_0}} \right) = {D_{KL}}\left( {q\left( {{{\bf{x}}_T}|{{\bf{x}}_0}} \right)\left\| {p\left( {{{\bf{x}}_T}} \right)} \right.} \right)$.

    \begin{figure*}[b]
	\normalsize
	\setcounter{mycount}{\value{equation}}
	\hrulefill
	\vspace*{4pt}
    \begin{align}\label{eq23}
    {L_\theta }\left( {{{\bf{x}}_0}} \right) = {\mathbb{E}_q}\left[ \begin{array}{l}
    {L_T}\left( {{{\bf{x}}_0}} \right) - \log {p_\theta }\left( {{{\bf{x}}_0}|{{\bf{x}}_1}} \right)
    + \sum\limits_{t > 1} {{D_{KL}}\left( {q\left( {{{\bf{x}}_{t - 1}}|{{\bf{x}}_t},{{\bf{x}}_0}} \right)\left\| {{p_\theta }\left( {{{\bf{x}}_{t - 1}}|{{\bf{x}}_t}} \right)} \right.} \right)}
    \end{array} \right]
    \end{align}
\end{figure*}
On this basis, we incorporate ${\bf{u}}$ as input for the reverse process to obtain
    \begin{align}\label{eq24}
    {p_\theta }\left( {{{\bf{x}}_{0:T}}\left| {\bf{u}} \right.} \right) = {p_\theta }\left( {{{\bf{x}}_T}} \right)\prod\limits_{t = 1}^T {{p_\theta }\left( {{{\bf{x}}_{t - 1}}|{{\bf{x}}_t},{\bf{u}}} \right)},
    \end{align}
    where
    \begin{align}\label{eq25}
    {p_\theta }\left( {{{\bf{x}}_{t - 1}}|{{\bf{x}}_t},{\bf{u}}} \right) = {{\cal N}}\left( {{{\bf{x}}_{t - 1}};{{\bm{\mu }}_\theta }\left( {{{\bf{x}}_t},t,{\bf{u}}} \right),{{\bm{\Sigma }}_\theta }\left( {{{\bf{x}}_t},t,{\bf{u}}} \right)} \right).
    \end{align}

    This converts~\ref{eq20} into a conditional model, thereby controlling the reverse diffusion process to generate the desired data. Meanwhile, in this paper, the data dimension is large and the valuable data points are scattered. For example, when $W=80$, a single V-A spectrum comprises 6561 data points. However, only some specific points, associated with moving human targets, carry the essential information, while the remaining ones do not offer much valuable information. Therefore, to guarantee that these important points play a more significant role during the training, we further introduce a matrix ${\bm{\psi }}$ to weight different points. \textcolor{black}{Specifically, ${\bm{\psi }}$ includes two weight factors, one is ${\psi_1=50}$, which is assigned to the data points that contain the vital information we need. The other one is ${\psi_0=1}$, which corresponds to the remaining points. Throughout the training, the values of these two parameters remain unchanged to ensure that important points can always receive more attention.} Therefore, based on the weighting matrix ${\bm{\psi }}$ and condition ${\bf{u}}$, the loss function is further optimized to
    \begin{equation}\label{eq26}
    {\mathbb{E}_{{{\bf{x}}_0},{\bm{\varepsilon }} \sim N\left( {0,{\bf{I}}} \right),t}}\left[ {{{\left\| {\left( \begin{array}{l}
    {{\bm{\varepsilon }}_\theta }\left( {\sqrt {{{\bar \alpha }_t}} {{\bf{x}}_0} + \sqrt {1 - {{\bar \alpha }_t}} {\bm{\varepsilon }},t} \right)\\
    - {\bm{\varepsilon }},t,{\bf{u}}
    \end{array} \right) \odot {\bm{\psi }}} \right\|}^2}} \right],
    \end{equation}

    where $ \odot $ is the Hadamard product operator.
     \begin{figure*}[htp]
    \centering
    \includegraphics[width=1\textwidth]{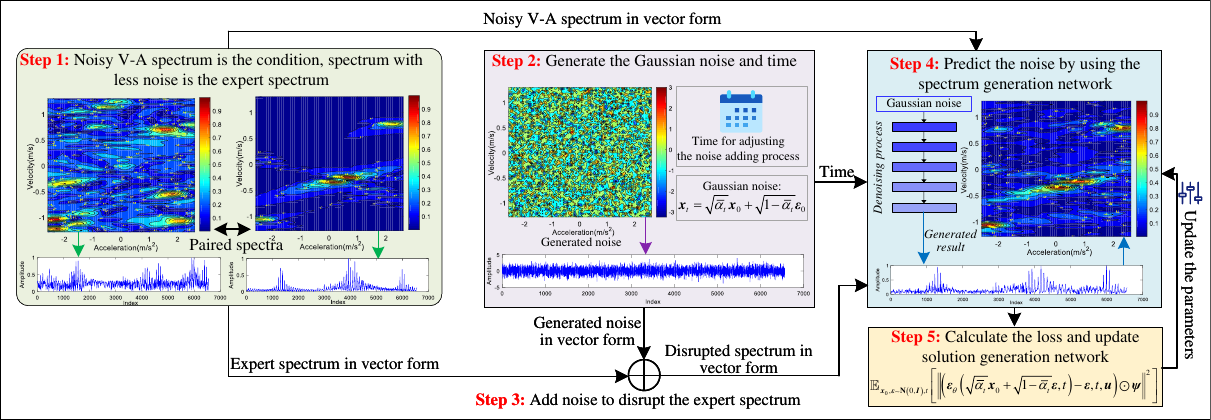}
    \caption{\textcolor{black}{The training process of the proposed UW-CDM for denoising V-A spectrum. First, the V-A spectrum with less noise serves as the expert spectrum, while the noisy V-A spectrum acts as the condition, and both are converted into vectors.Then, Step 2 generates the Gaussian noise with intensity controlled by network hyperparameters and the steps for adding noise. Based on this, the Gaussian noise is added to the expert spectrum as the step requirements, thereby disrupting the expert spectrum, as illustrated in Step 3. Subsequently, the contaminated expert spectrum is fed to Step 4 for denoising, which involves learning how to reduce noise at each step to generate the spectrum. Here, the goal of the loss function, which is based on denoising score matching, is to reduce the gap between the added noise that disrupts the expert spectrum and the noise estimated by the model, while considering the step $t$ and the current condition $\bf{u}$. After training, the UW-CDM is capable of generating the V-A spectrum with less noise using the spectrum obtained from velocity and acceleration estimation, thereby achieving the denoising.}}
    \label{CDM}
    \vspace{-0.5cm}
    \end{figure*}

    Based on the aforementioned principles, we use the noisy V-A spectrum as $\bf{u}$ and the denoised V-A spectrum as the generation target to train the UW-CDM. Specifically, the model training process is illustrated in Fig.~\ref{CDM}, which includes five steps:

    \begin{enumerate}
    \item The first step is to obtain the paired V-A spectra with significant noise and less noise, and then convert them into the vectors\footnote{Paired spectra are generated via simulation, with target counts, velocities, and accelerations assigned randomly. Target counts range from 1 to 7, velocities span -2 to 2 m/s, and accelerations from -3 to 3 m/s². Noise spectra are produced at a signal-to-noise ratio (SNR) of -10 dB , while expert spectra are generated at an SNR of 10 dB.}. During the training process, the noisy vector is used as a condition (i.e., $\bf{u}$ in the UW-CDM), while the spectrum with less noise is used as the expert spectrum, i.e., the one we want UW-CDM to produce during the data generation process.

    \item The second step is to generate the Gaussian noise and time. This time will be used to adjust the process of adding noise to the data.

    \item The third step involves adding the generated noise to the expert spectrum, so as to disturb it.

    \item The fourth step is to feed the noisy V-A spectrum, generated time, and the disturbed expert spectrum into the diffusion model to train the denoising network.

    \item Finally, we compute the loss based on (\ref{eq26}) and update the parameters of the denoising network according to the results.

    \end{enumerate}

    Once trained, the UW-CDM can generate the V-A spectrum with less noise based on the given condition, i.e., the noisy V-A spectrum obtained by the velocity and acceleration estimation, thereby achieving V-A spectrum denoising. In the generated V-A spectrum, the peak points corresponding to the moving human targets become more distinct and are easier to spot. Therefore, by setting a threshold, the number of peak points and their coordinates can be identified, enabling us to detect the number of moving human targets and obtain their velocities and accelerations.

    \subsection{DoA and ToF Estimation}
    After obtaining the number of moving human targets, the next step is to determine the number of subflows and the size of each subflow. As illustrated in Fig.~\ref{application}, a subflow is composed of one or more human targets, which are closely spaced and moving toward the same direction with similar walking velocities. Therefore, this section estimates the DoA and ToF of HIR, which unveil the spatial information of moving human targets, crucial for subsequent human flow detection.
    \subsubsection{DoA Estimation}
    In practical communication scenarios, the antenna spacing in the ULA might exceed half a wavelength. Hence, in G-HFD, we consider that the antenna spacing of the ULA is one wavelength (i.e., $\lambda $) and estimate the DoA based on the obtained CSI. Specifically, equation (\ref{eq6}) gives the CSI of the cross component at time $\Delta t$ from the $m$-th antenna. Then, the cross component of the CSI from the $m+1$-th antenna is
    \begin{align}\label{eq27}
    {{h'}_{\left( {m + 1} \right)\bar r,k}}\left( {\Delta t} \right) &= \sum\limits_{r' = 1}^{R'} {\left( {\alpha _{\left( {m + 1} \right)\bar r,k}^{\left[ {r'} \right]}} \right.{e^{ - j2\pi {f_k}\left[ {\tau _{\left( {m + 1} \right)\bar r}^{\left[ {r'} \right]} + \frac{{d\sin \left( {{\varphi _{r'}}} \right)}}{c}} \right]}}}\notag \\
    &\times \left. {{e^{ - j2\pi {f_k}\left[ {\frac{{{v_{r'}}\Delta t}}{c} + \frac{{{a_{r'}}{{\left( {\Delta t} \right)}^2}}}{{2c}}} \right]}}} \right),
    \end{align}
    where $d=\lambda$ is the antenna spacing and ${\varphi _{r'}}$ is the signal DoA. From (\ref{eq6}), (\ref{eq7}), and ${{\bf{S}}_{m,k}}$, we can see that after eliminating the interference, (\ref{eq27}) is included in ${s_{m + 1,k}}\left( {\Delta t} \right)$. When $d \le {\lambda  \mathord{\left/  {\vphantom {\lambda  2}} \right.  \kern-\nulldelimiterspace} 2}$, the MUSIC algorithm~\cite{schmidt1986multiple} can effectively estimate the signal DoA based on $\left[ {{s_{1,k}}\left( {\Delta t} \right),{s_{2,k}}\left( {\Delta t} \right),{\rm{ }} \ldots {\rm{, }}{s_{M,k}}\left( {\Delta t} \right)} \right]$. However, when $d > {\lambda  \mathord{\left/  {\vphantom {\lambda  2}} \right.  \kern-\nulldelimiterspace} 2}$, the estimated DoA spectrum becomes ambiguous\footnote{When a single signal arrives at the ULA, the antenna spacing exceeding half the wavelength can result in multiple peak points in the spectrum obtained via DoA estimation. This effect, caused by the periodicity of the phase, prevents the system from identifying the true DoA of the signal from the estimated results.} due to the periodicity of the phase.
    % \begin{figure*}[htp]
    % \centering
    % \includegraphics[width=1\textwidth]{Image/AMBv3.pdf}
    % \caption{The cause of ambiguous DoA spectrum.}
    % \label{AMB}
    % \end{figure*}

    To accurately extract the true DoA of HIR, G-HFD needs to eliminate ambiguous peaks. At the same time, for effective subsequent subflow analysis, the estimated DoA must be associated with the obtained velocity. Hence, we first use the obtained velocity and acceleration to rotate the phase of CSI and perform joint velocity and DoA estimation to generate the ambiguous DoA spectrum. Then, the proposed UW-CDM method is used to generate the clear spectrum by using the ambiguous spectrum as the condition, thereby realizing the DoA estimation when $d = \lambda $. Concretely, $W$ CSI observations from $M$ antennas are used to build
    \begin{align}\label{eq28}
    {{\bf{S}}_{M - W}} = \left[ \begin{array}{l}
    {s_{1,k}}\left( 0 \right),{\rm{ }} \ldots {\rm{, }}{\ }{s_{1,k}}\left( {\left( {W - 1} \right)\Delta t} \right)\\
    {\qquad\qquad\quad\quad} \vdots \\
    {s_{M,k}}\left( 0 \right),{\rm{ }} \ldots {\rm{, }}{\ }{s_{M,k}}\left( {\left( {W - 1} \right)\Delta t} \right)
    \end{array} \right].
    \end{align}
    Then, based on (\ref{eq28}) and the estimated velocity and acceleration, the rotation matrix is constructed as
    \begin{align}\label{eq29}
    {{\bf{P}}_{M - W}} = \left[ \begin{array}{l}
    1,{\rm{ }} \ldots {\rm{, }}{\ }{e^{j2\pi {f_k}\left[ {\left( {W - 1} \right)\left( {\frac{{{{\hat v}_{r'}}\Delta t}}{c} + \frac{{{{\hat a}_{r'}}{{\left( {\Delta t} \right)}^2}}}{{2c}}} \right)} \right]}}\\
    {\qquad\qquad\qquad\quad} \vdots \\
    1,{\rm{ }} \ldots {\rm{, }}{\ }{e^{j2\pi {f_k}\left[ {\left( {W - 1} \right)\left( {\frac{{{{\hat v}_{r'}}\Delta t}}{c} + \frac{{{{\hat a}_{r'}}{{\left( {\Delta t} \right)}^2}}}{{2c}}} \right)} \right]}}
    \end{array} \right],
    \end{align}
    where ${\hat v_{r'}}$ and ${\hat a_{r'}}$ are the estimated velocity and acceleration corresponding to the $r'$-th HIR, respectively. After that, the phase rotation is realized by multiplying (\ref{eq28}) with (\ref{eq29}) to obtain ${{\bf{S'}}_{M - W}} = {{\bf{S}}_{M - W}} \odot {{\bf{P}}_{M - W}}$. This multiplication operation eliminates the phase accumulation caused by ${\hat v_{r'}}$ and ${\hat a_{r'}}$, so that in ${{\bf{S'}}_{M - W}}$ the velocity and acceleration of $r'$-th HIR are zero.

    After that, the two-dimensional MUSIC (2D-MUSIC) algorithm is performed on ${{\bf{S'}}_{M - W}}$ for joint velocity and DoA estimation. As the velocity and acceleration corresponding to $r'$-th HIR have been rotated to zero, during the parameter search process, we fix the velocity at zero and search for the DoA within the range of -90 degrees to 90 degrees to achieve the estimation. Figure~\ref{2D1D} gives a set of simulation estimation results, including those before and after phase rotation. At the same time, the estimation results when $d = {\lambda/2}$ are also provided for reference. From Fig.~\ref{2D1D}, we can see that:
    \begin{figure*}[htp]
    \centering
    \includegraphics[width=1\textwidth]{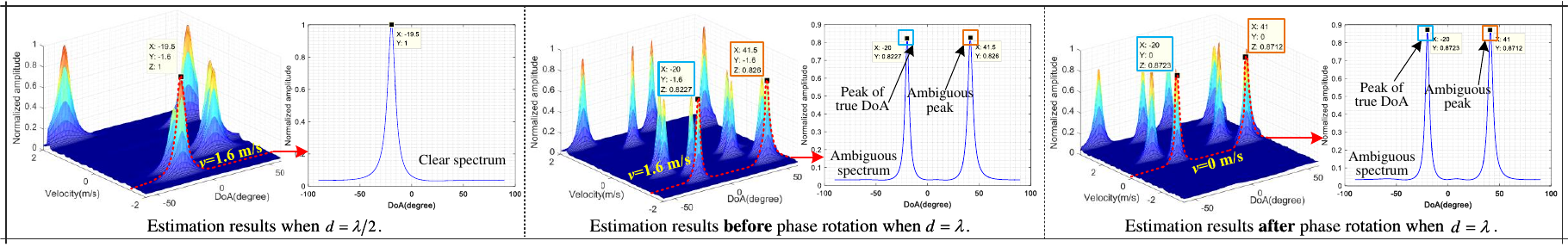}
    \caption{The joint velocity and DoA estimation results under different cases when there are six reflections.}
    \label{2D1D}
    \vspace{-0.3cm}
    \end{figure*}

    % \begin{figure}[t]
    % \centering
    % \includegraphics[height=10cm]{Image/2D1DV4.pdf}
    % \caption{The joint velocity and DoA estimation results under different cases when there are six reflections.}
    % \label{2D1D}
    % \vspace{-0.5cm}
    % \end{figure}

    \begin{enumerate}
    \item Compared with the result when $d = {\lambda/2}$, the joint estimation yields an ambiguous DoA spectrum when $d=\lambda$, that is, a single HIR triggers two peaks with different DoAs.

    \item After phase rotation, the velocity corresponding to the reflection changes from 1.6 $m/s$ to 0 $m/s$. This allows for the DoA search by fixing the velocity to 0 $m/s$ during the estimation process, which not only reduces the computational load but also matches the estimated velocity with the DoA.
    \end{enumerate}

    Based on the ambiguous DoA spectrum, the proposed UW-CDM is employed to generate the true DoA. We train the UW-CDM with paired clear and ambiguous DoA spectra, which are obtained by simulation under the condition of $d = {\lambda/2}$ and $d = \lambda $, respectively. The training process is the same as that presented in Fig.~\ref{CDM}. However, here, the expert spectrum and condition $\bf{u}$ are the clear DoA spectrum and ambiguous DoA spectrum, respectively. After training, the UW-CDM can generate the clear DoA spectrum based on the provided ambiguous one, thereby achieving DoA estimation. The overall estimation process is summarized in Algorithm 1.

\subsubsection{ToF Estimation}
Similar to DoA, we use the phase rotation combined with joint estimation to obtain the ToF of each HIR. Specifically, the $W$ CSI
observations from $K$ subcarriers, ${\hat v_{r'}}$, and ${\hat a_{r'}}$ are used to construct matrices
\begin{align}\label{eq30}
{{\bf{S}}_{K - W}} = \left[ \begin{array}{l}
{s_{m,1}}\left( 0 \right),{\rm{ }} \ldots {\rm{, }}{s_{m,1}}\left( {\left( {W - 1} \right)\Delta t} \right)\\
{{\qquad\qquad\qquad\qquad}} \vdots \\
{s_{m,K}}\left( 0 \right),{\rm{ }} \ldots {\rm{, }}{s_{m,K}}\left( {\left( {W - 1} \right)\Delta t} \right)
\end{array} \right]
    \end{align}
and
\begin{align}\label{eq31}
{{\bf{P}}_{K - W}} = \left[ \begin{array}{l}
1,{\rm{ }} \ldots {\rm{, }}{e^{j2\pi {f_1}\left[ {\left( {W - 1} \right)\left( {\frac{{{{\hat v}_{r'}}\Delta t}}{c} + \frac{{{{\hat a}_{r'}}{{\left( {\Delta t} \right)}^2}}}{{2c}}} \right)} \right]}}\\
{\qquad\qquad\qquad\qquad} \vdots \\
1,{\rm{ }} \ldots {\rm{, }}{e^{j2\pi {f_K}\left[ {\left( {W - 1} \right)\left( {\frac{{{{\hat v}_{r'}}\Delta t}}{c} + \frac{{{{\hat a}_{r'}}{{\left( {\Delta t} \right)}^2}}}{{2c}}} \right)} \right]}}
\end{array} \right],
\end{align}
where ${f_K} = {f_1} + (K - 1)\Delta f$, $\Delta f$ is the frequency interval between two adjacent subcarriers, and $f_1$ is the minimum frequency. Then, multiplying (\ref{eq30}) and (\ref{eq31}) results in ${{\bf{S'}}_{K - W}} = {{\bf{S}}_{K - W}} \odot {{\bf{P}}_{K - W}}$. On this basis, using the 2D-MUSIC algorithm, the ToF of each HIR can be estimated. Unlike the DoA estimation, here the search range of ToF during the estimation is determined by $\Delta f$, thereby guaranteeing a clear ToF spectrum and effective estimation.
 \begin{algorithm}[t]
{\small \caption{DoA Estimation When Antenna Spacing $d=\lambda$}}
% \hspace*{0.02in} {\bf{\textit{Training Phase:}}}
\begin{algorithmic}[1]
% \State Input hyper-parameters: denoising step $T$, initialize neural network parameters $\omega $ and $v$
% \vspace{0.1cm}
% \State \#\#{\textit{\quad Learning Process}}
% \State Initialize a random process for pricing strategy exploration
\For {$r' = 1:R'$}
\State Construct the ${{\bf{P}}_{M - W}}$ based on estimated ${\hat v_{r'}}$ and ${\hat a_{r'}}$
\State Phase rotation via ${{\bf{S'}}_{M - W}} = {{\bf{S}}_{M - W}} \odot {{\bf{P}}_{M - W}}$
\State Calculate the auto-correlation matrix of ${{\bf{S'}}_{M - W}}$ and perform singular value decomposition to obtain the noise vector
\State Construct the steering matrix based on (\ref{eq27}) and (\ref{eq28}), and combine it with the noise vector to formulate the spectrum function
\State Fix $v=0$ and search for DoA within the range of -90 to 90 degrees to produce the ambiguous DoA spectrum
\State Use the ambiguous spectrum as condition $\bf{u}$ and employ the trained UW-CDM to produce the corresponding clear spectrum
\State Extract the DoA of $r'$-th signal from the clear spectrum
\EndFor
\State \Return DoAs of $R'$ signals
\end{algorithmic}
\end{algorithm}
\subsection{Flow Detection}
Based on the aforementioned calculations, we obtain the total number of moving human targets, along with their velocities, DoAs, and ToFs. Given that human targets within the same subflow are close and maintain similar velocities, their corresponding parameters tend to cluster together. Therefore, this section conducts two-stage clustering to determine the number of subflows and the subflow size.

\begin{figure}[t]
\centering
\includegraphics[height=1.5cm]{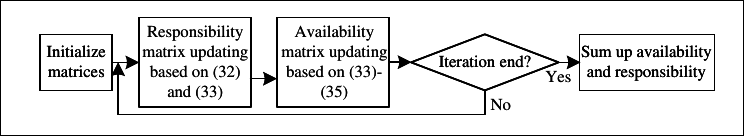}
\caption{The affinity propagation clustering process.}
\label{CLUST}
\vspace{-0.5cm}
\end{figure}

Specifically, we first use the parameters obtained from multiple calculations to build a set $Y = \left( {{{\bf{y}}_i}\left| {i = 1, \ldots, I} \right.} \right)$, where $I$ is the total number of data points, ${{\bf{y}}_i} = \left( {{{\hat v}_i},{{\hat \varphi }_i},{{\hat \tau }_i}} \right)$, ${\hat v_i}$, ${\hat \varphi _i}$, and ${\hat \tau _i}$ are the estimated velocity, DoA, and ToF, respectively. Then, in the first stage, the adaptive affinity propagation clustering~\cite{wang2008adaptive} is conducted on $Y$ to determine the total number of subflows. The key idea in affinity propagation clustering is to transmit and iterate responsibility and availability among the data points in $Y$ to generate several cluster centers. On this basis, the rest of data points are assigned to the respective cluster based on these centers, thereby completing the clustering. The overall flow is given in Fig.~\ref{CLUST}. During this process, the similarity between the two data points is defined as
\begin{align}\label{eq32}
sim\left( {{{\bf{y}}_i},{{\bf{y}}_{i'}}} \right) =  - {\left\| {{{\bf{y}}_i} - {{\bf{y}}_{i'}}} \right\|_2},
\end{align}
where ${\left\|  \cdot  \right\|_2}$ is the second norm operator, and the responsibility can be calculated as
\begin{align}\label{eq33}
res\left( {{{\bf{y}}_i},{{\bf{y}}_{i'}}} \right)& = sim\left( {{{\bf{y}}_i},{{\bf{y}}_{i'}}} \right)\notag \\ & - \mathop {\max }\limits_{i'' \ne i'} \left\{ {ava\left( {{{\bf{y}}_i},{{\bf{y}}_{i''}}} \right) + sim\left( {{{\bf{y}}_i},{{\bf{y}}_{i''}}} \right)} \right\}.
\end{align}
Meanwhile, the availability used in (33) is defined as
\begin{align}\label{eq34}
ava\left( {{{\bf{y}}_i},{{\bf{y}}_{i'}}} \right) = \min \left\{ \begin{array}{l}
0,{\rm{ }}res\left( {{{\bf{y}}_{i'}},{{\bf{y}}_{i'}}} \right)\\
 + \sum\limits_{i''' \notin \left\{ {i,i'} \right\}} {\max \left\{ {0,res\left( {{{\bf{y}}_{i'''}},{{\bf{y}}_{i'}}} \right)} \right\}}
\end{array} \right\},
\end{align}
when $i \ne i'$. Otherwise we have
\begin{align}\label{eq35}
ava\left( {{{\bf{y}}_i},{{\bf{y}}_{i'}}} \right) = \sum\limits_{i''' \ne i'} {\max \left\{ {0,{\rm{ }}res\left( {{{\bf{y}}_{i'''}},{{\bf{y}}_{i'}}} \right)} \right\}}.
\end{align}

Here the $res\left( {{{\bf{y}}_i},{{\bf{y}}_{i'}}} \right)$ indicates the accumulated evidence that sample ${{\bf{y}}_{i'}}$ is the cluster center of sample ${{\bf{y}}_{i}}$, while $ava\left( {{{\bf{y}}_i},{{\bf{y}}_{i'}}} \right)$ represents the accumulated evidence that sample ${{\bf{y}}_{i}}$ selects sample ${{\bf{y}}_{i'}}$ as its cluster center. After several iterations, if the selected cluster centers remain unchanged, then we sum up the responsibility and availability and output the clustering results. These results include the number of clusters and the data points contained in each cluster, where the number of clusters represents the number of subflows.

\begin{figure*}[htp]
    \centering
    \includegraphics[width=0.95\textwidth]{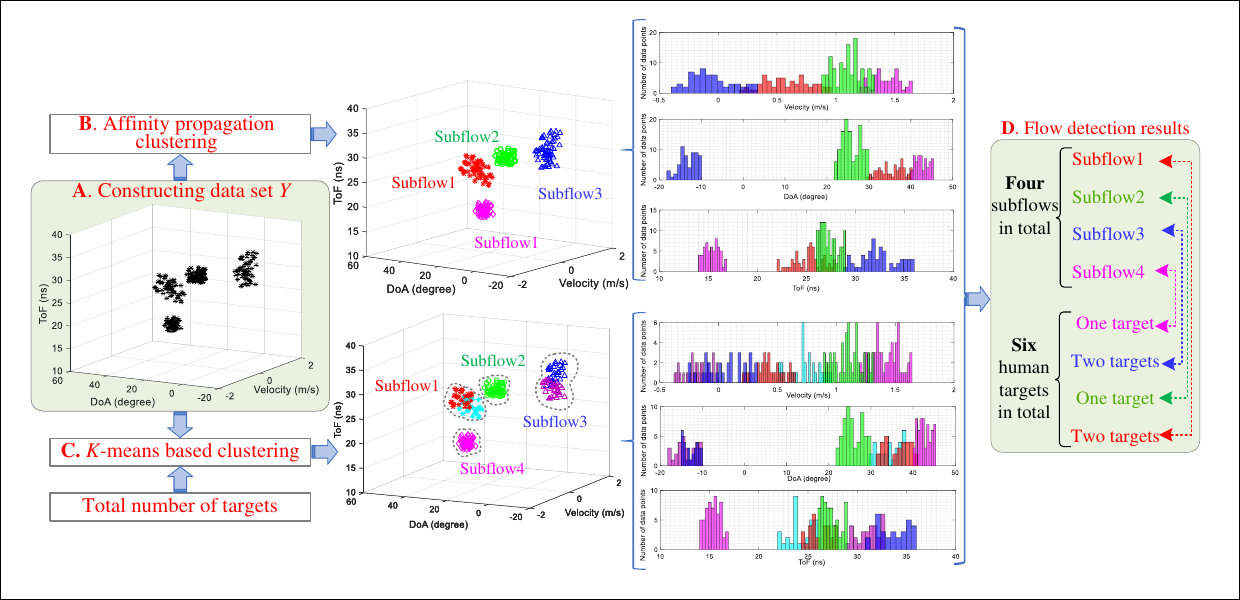}
    \caption{The flow detection based on clustering. G-HFD first performs the affinity propagation clustering to detect the number of subflows. Then, using the total number of human targets and subflows, G-HFD performs the $K$-means based clustering to identify the size of each subflow.}
    \label{PITS}
    \end{figure*}

After determining the number of subflows, G-HFD further needs to identify how many moving targets are within each subflow. Therefore, in the second stage, we perform $K$-means clustering on $Y$ to divide it into $\Upsilon$ clusters, where $\Upsilon$ is the obtained total number of human targets. The clustering process is presented in Algorithm 2. After that, by comparing the outcomes of the two-stage clustering, we can determine the number of clusters that are covered by the data points in each subflow, therefore determining the number of human targets contained in each subflow. It is worth noting that data points corresponding to a single human target may span across multiple subflows. In such instances, the target is finally assigned to the subflow that contains the majority of the data points. Figure~\ref{PITS} shows the flow detection process based on the data collected from real world scenario.
\vspace{-0.3cm}

   \section{Implementation and Evaluation}
    In this section, we build a platform based on the commercial routers and conduct tests to evaluate the proposed system in practical communication scenarios. The evaluation mainly includes three aspects: (i) the effectiveness of the proposed UW-CDM network in V-A spectrum denoising and DoA spectrum generation; (ii) the performance in detecting the number of human targets; and (iii) the performance in detecting the number of subflows and the size of each subflow.
    \vspace{-0.5cm}
\subsection{Experimental Configurations}
\subsubsection{\textbf{Experimental Scenarios}}
\textcolor{black}{We choose a representative corridor and a meeting room as our experimental scenario, depicted in Fig.~\ref{HDW2}. This corridor is characterized by two elevator exits, one emergency exit, and two intersections connecting to two separate corridors. Compared with corridors, the conference room is more enclosed and contains many tables and chairs, resulting in a greater number of reflected signals. Hence, the conference room presents a more complex environment than the corridor. In typical office environment, these areas often experience dense pedestrian movement with unpredictable patterns, marking them the key areas for flow monitoring.}
\subsubsection{\textbf{Hardware Configuration}}
\begin{figure*}[htp]
    \centering
    \includegraphics[width=1\textwidth]{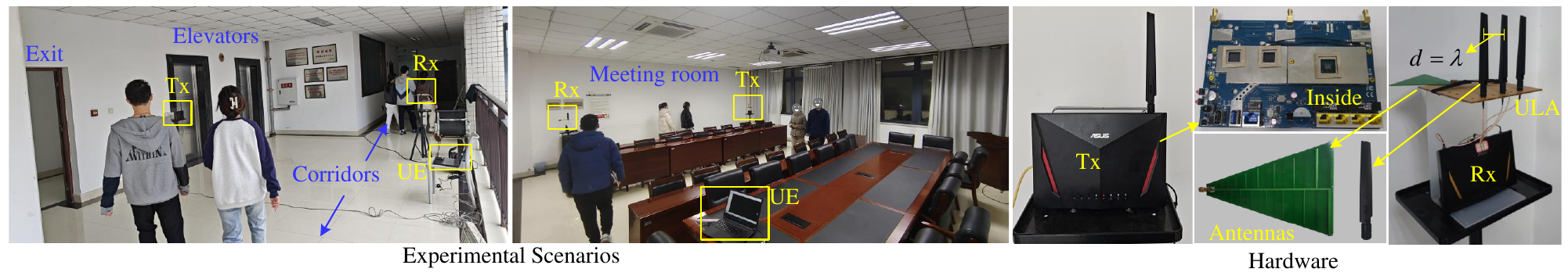}
    \caption{The experimental scenarios and hardware configurations.}
    \label{HDW2}
    \end{figure*}

In this paper, two commercial ASUS APs and one UE are used to complete the experiments, as shown in Fig.~\ref{HDW2}. Specifically, one AP acts as Tx to provide network services, while the other one, equipped with the Nexmon toolbox, acts as the Rx to capture signals and extract CSI. This Rx has four radio-frequency channels, one with a directional antenna for direct signal reception from the Tx, and the other three form a ULA with the antenna spacing of one wavelength to capture multipath signals. The Tx uses an omnidirectional antenna to offer Internet services. Throughout the experiment, the signal frequency is fixed at 5.805 GHz, with a bandwidth of 80 MHz.
   \begin{algorithm}[t]
{\small \caption{Clustering for Subflow Size Detection}}
\begin{algorithmic}[1]
\State Input the set $Y$ and parameter $\Upsilon $ as the number of classes.
\State Find out ${Y_p}$ that has the minimum sum of the squared error (SSE) in $Y$
\State Utilize the $K$-means algorithm ($K=2$) to perform $\Phi $ times of classification on the class ${Y_p}$ to obtain $C = \left\{ {{C_1}, \ldots ,{C_\Phi }} \right\}$, where ${C_\phi }{\rm{ = }}\left\{ {{Y_{\phi 1}},{Y_{\phi 2}}} \right\}$
\State Single out the class ${C_\phi }{\rm{ = }}\left\{ {{Y_{\phi 1}},{Y_{\phi 2}}} \right\}$, which has the minimum SSE
\State Add ${C_\phi }{\rm{ = }}\left\{ {{Y_{\phi 1}},{Y_{\phi 2}}} \right\}$ to $Y$ and remove ${Y_p}$ from $Y$
\State Repeat the above steps until $\Upsilon $ clusters are obtained
\State \Return cluster label of each data point
\end{algorithmic}
\end{algorithm}
\subsubsection{\textbf{Experimental Method}}
During the experiments, the UE connects to the Tx and performs three Internet activities, including downloading files, online games, and watching videos. According to the activities, the Tx transmits the data to the UE with varying packet transmission rates. In these three cases, the Rx captures the downstream signals to evaluate the system's performance without affecting the communication between the UE and Tx. In the experiment of detecting the total number of human targets, a variable number (up to ten) of human targets randomly form subflows and walk through the monitored area to facilitate data collection. For the subflow detection, ten human targets randomly form some subflows and walk through the detection area for data collection. All the gathered data is then processed offline on a server, which is installed with the Ubuntu 20.04 operating system and equipped with an AMD Ryzen Threadripper PRO 3975WX 32-core processor and an NVIDIA RTX A5000 graphics processing unit (GPU). It is worth noting that we train the proposed UW-CDM network by using the paired data generated through simulations. Subsequently, the trained model is employed to process the collected data to complete the evaluation.
\subsubsection{\textbf{Evaluation Metrics}}
We employed detection accuracy (DA) as the metric for assessing the system's performance in human flow detection. This metric is defined as the ratio of the number of correct detections to the total number of experiments conducted. Taking the detection of the number of subflows as an example, if out of 100 experiments, G-HFD correctly detects 90 times, then the DA is 90\%. For subflow size detection, a correct detection means that the number of human targets in each subflow is accurately detected. Meanwhile, we also use the confusion matrix to analyze the flow detection performance under different conditions.

\vspace{-0.3cm}
\subsection{Experimental Results}
\subsubsection{\textbf{V-A Spectrum Denoising}}
\begin{figure*}[htbp]
%\vspace{-.1cm}
\centering
%\hspace{-.5cm}
\subfigure[The input noisy V-A spectrum.]{
\begin{minipage}[t]{0.2425\linewidth}
\centering
\includegraphics[width=4.4cm]{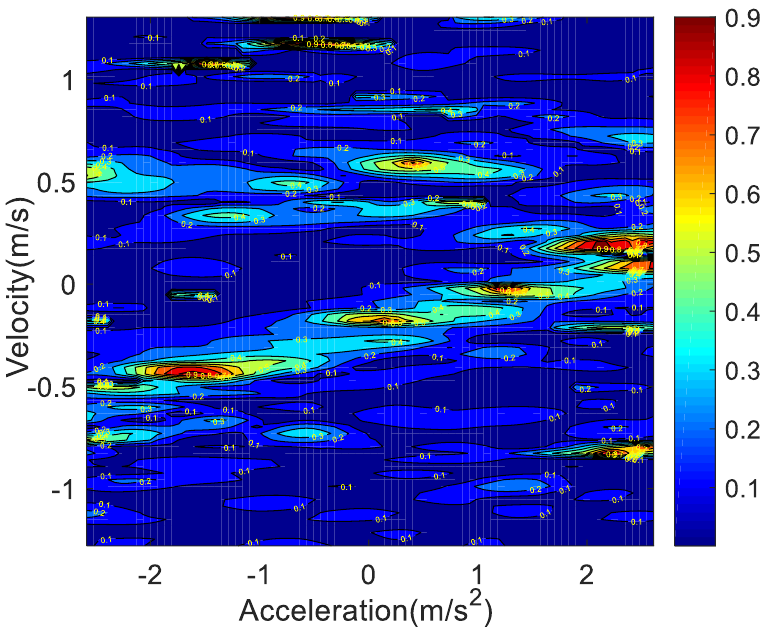}
%\caption{fig1}
\end{minipage}%
}%
\subfigure[The denoising result of UW-CDM.]{
\begin{minipage}[t]{0.2415\linewidth}
\centering
\includegraphics[width=4.5cm]{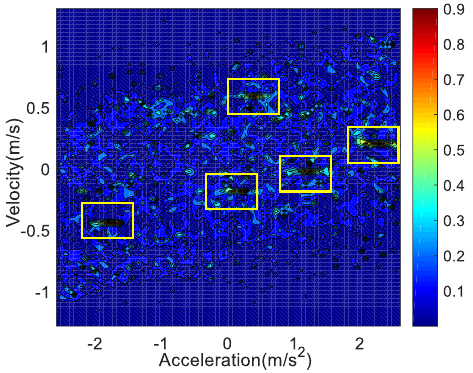}
%\caption{fig2}
\end{minipage}%
}%
\subfigure[The denoising result of PMRID.]{
\begin{minipage}[t]{0.2345\linewidth}
\centering
\includegraphics[width=4.5cm]{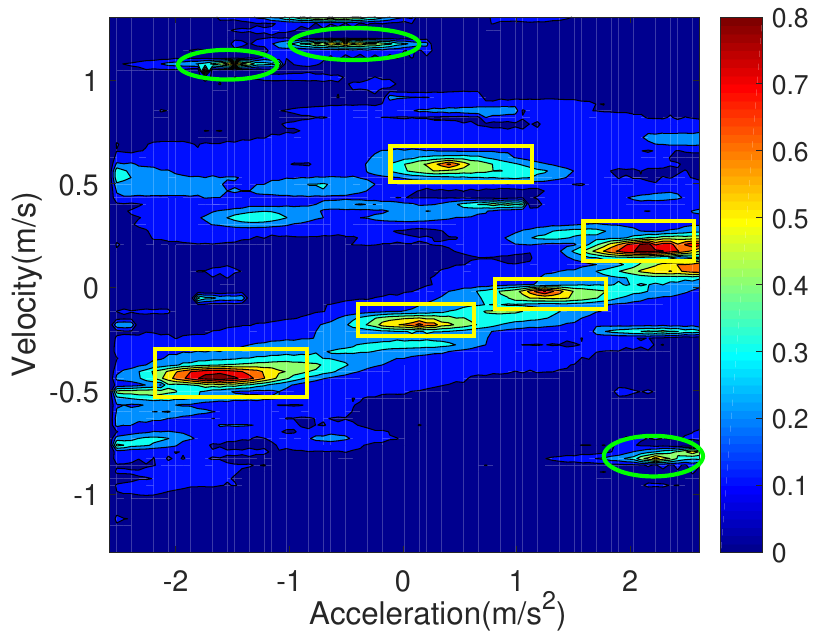}
%\caption{fig2}
\end{minipage}
}%
\subfigure[The denoising result of DPIR.]{
\begin{minipage}[t]{0.245\linewidth}
\centering
\includegraphics[width=4.5cm]{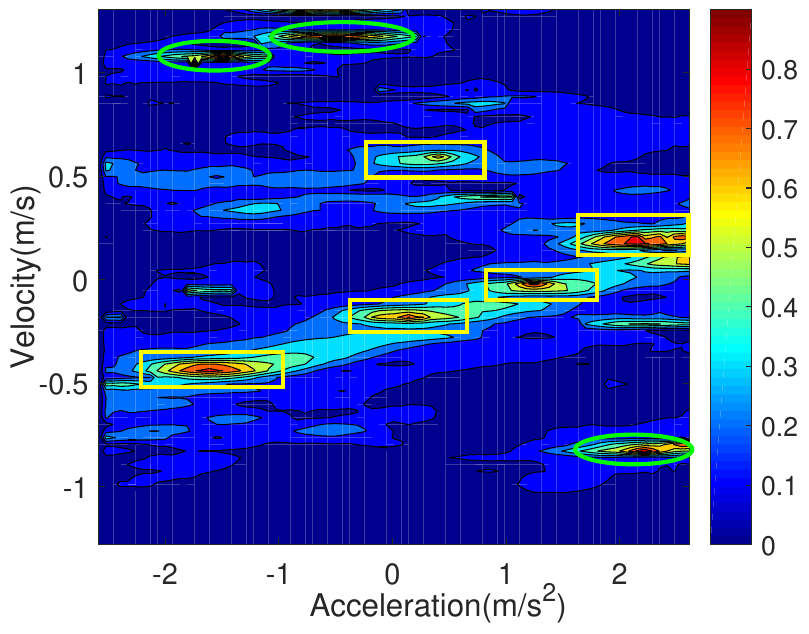}
%\caption{fig2}
\end{minipage}
}%
\caption{\textcolor{black}{The denoising performance comparison of different models.}}
\label{VAD}
\end{figure*}

 \textcolor{black}{We first assess the denoising performance of UW-CDM in a practical communication scenario where the UE is downloading files, and compare it with two CNN-based denoising models proposed in~\cite{wang2020practical} and~\cite{zhang2021plug}, which are denoted as PMRID and DPIR, respectively.} Using five human targets walking at a normal speed as examples, Figs.~\ref{VAD}(a) and (b) present the input noisy spectrum and the output of UW-CDM, respectively. As can be seen, after ten inference steps, UW-CDM can effectively produce the denoised V-A spectrum. Compared with the input noisy V-A spectrum, two observations stand out. First, without denoising, identifying the peak points corresponding to the targets is challenging due to the interference introduced by signal noise, cross terms, etc. After denoising, the peak points corresponding to the human targets become more discernible, which facilitates a straightforward and accurate detection of the number of moving human targets. Second, after the denoising, the positions of the peak points corresponding to the human targets are nearly the same as before denoising. This consistency indicates that the denoising does not affect the coordinates (i.e., the acceleration and velocity) of the peak points in the spectrum, verifying the effectiveness of UW-CDM in V-A spectrum denoising. \textcolor{black}{Note that with our platform, ten inference steps require about 2.1 seconds, which may not be able to support applications that require low response delay. Fortunately, many methods were proposed to accelerate the inference in diffusion models~\cite{chen2023speed}, which can significantly reduce the response delay of the proposed model, thereby enabling real-time applications.}

 \textcolor{black}{Meanwhile, as depicted in Figs.~\ref{VAD}(c) and (d), PMRID and DPIR also demonstrate denoising capabilities while keeping the positions of the target peaks, as indicated by the boxes in the figure. However, compared with UW-CDM, their denoising is not as thorough. Specifically, in the denoised V-A spectrum, besides the peaks corresponding to the targets, there are still some other noise peaks with relatively strong energy, as marked by circles in Figs.~\ref{VAD}(c) and (d). These noise peaks could be mistakenly recognized as target peaks and hence affect the detection accuracy. Therefore, the denoising performance of the proposed diffusion model-based UW-CDM is superior to that of the CNN-based PMRID and DPIR, offering more thorough noise cancellation and ensuring the detection performance of the number of targets.}

\begin{figure*}[htbp]
\subfigure[ ]{
\begin{minipage}[t]{0.23\linewidth}
% \vspace{-.1cm}
% \centering
\includegraphics[width=4.5cm]{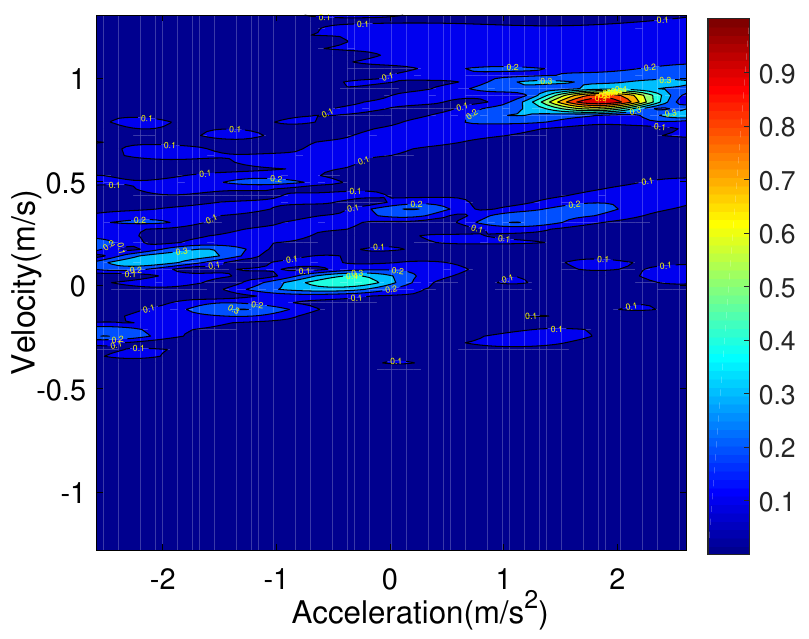}
%\caption{fig2}
\end{minipage}
}%
\subfigure[ ]{
\begin{minipage}[t]{0.235\linewidth}
\centering
\includegraphics[width=4.5cm]{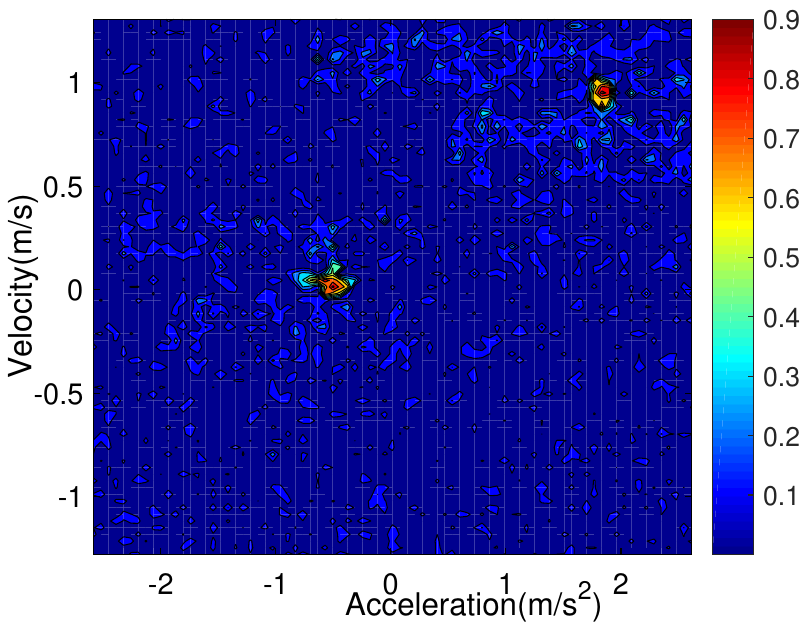}
%\caption{fig2}
\end{minipage}
}%
\subfigure[ ]{
\begin{minipage}[t]{0.23\linewidth}
\centering
\includegraphics[width=4.5cm]{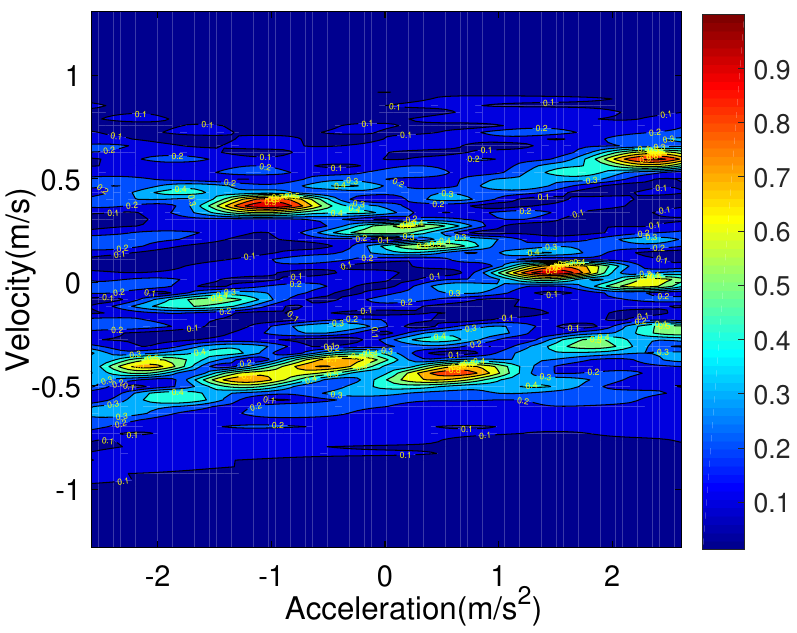}
%\caption{fig2}
\end{minipage}
}
\subfigure[ ]{
\begin{minipage}[t]{0.23\linewidth}
\centering
\includegraphics[width=4.5cm]{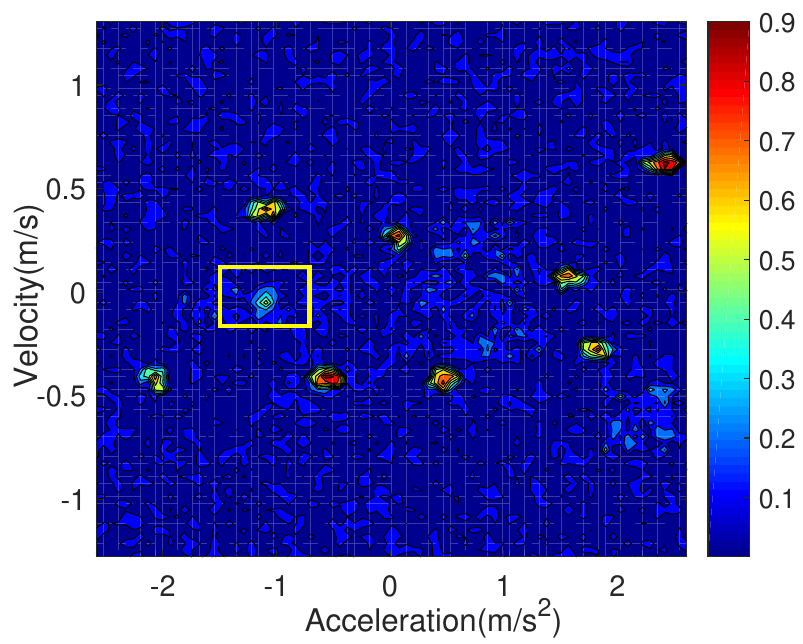}
%\caption{fig2}
\end{minipage}
}
\caption{\textcolor{black}{The denoising results at various noise intensities. (a) and (c) are the noisy V-A spectra when there are two and eight targets, respectively, and (b) and (d) are the corresponding denoising results.}}
\label{NSITS}
\end{figure*}
 \textcolor{black}{Upon validating the effectiveness, we analyze its denoising performance when a different number of targets appear, i.e., under varying noise intensities, with the results provided in Fig.~\ref{NSITS}. As can be seen, with two targets, UW-CDM effectively eliminates noise in the V-A spectrum so that the peaks of the targets can be easily and accurately detected. As the number of targets increases to seven, the interaction among targets intensifies, causing an increase in noise within the V-A spectrum, as shown in Fig.~\ref{NSITS}(c). Under these conditions, the performance of UW-CDM exhibits a slight decline, evidenced by the incomplete cancellation of some noise points, as marked in Fig.~\ref{NSITS}(d). However, overall, UW-CDM remains effective in eliminating strong noise peaks, making the peaks corresponding to targets more pronounced and thereby ensuring their efficient and accurate detection.}

\subsubsection{\textbf{DoA Spectrum Generation}}
\begin{figure*}[htp]
    \centering
    \includegraphics[width=1\textwidth]{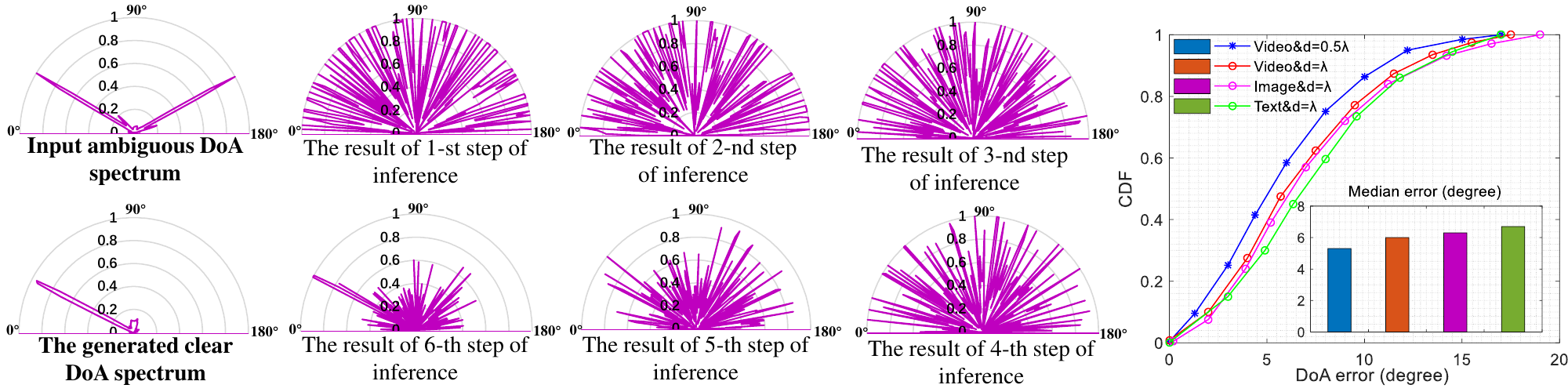}
    \caption{The generation process of the clear DoA spectrum based on the trained UW-CDM. }
    \label{DoAGn}
    \vspace{-0.5cm}
    \end{figure*}

Following V-A spectrum denoising, we evaluate UW-CDM's performance in generating the clear DoA spectrum. During the tests, a single human target walks along predetermined routes while the UE downloads files, so that the ground truth for DoA can be obtained for accuracy analysis. Figure~\ref{DoAGn} shows the DoA spectrum generation process. From the figures, we can see that, starting with Gaussian noise, the trained UW-CDM can effectively generate the clear DoA spectrum based on the given ambiguous DoA spectrum, by leveraging the reverse diffusion process. Unlike the input ambiguous spectrum, the generated one contains only one peak point corresponding to the single moving human target, enabling the proposed G-HFD to extract the DoA of the HIR when the antenna spacing of ULA is $\lambda$.

After that, we analyze the accuracy of DoA in the generated spectrum when the UE performs different activities, and compare it with the DoA estimation accuracy when the antenna spacing is half a wavelength (i.e., $d=0.5 \lambda$). The cumulative distribution function (CDF) presented in Fig.~\ref{DoAGn} reveals that the median DoA estimation error is about 5.2 degrees when the UE downloads files and $d=0.5 \lambda$. Meanwhile, the median DoA errors of UW-CDM are 6.1, 6.3, and 6.7 degrees, when UE performs downloading files, online games, and watching videos, respectively, which are comparable to the accuracy obtained when $d=0.5 \lambda$. These results further demonstrate the effectiveness of UW-CDM in clear DoA spectrum generation when $d= \lambda$, providing reliable parameter support for subsequent human flow detection.

\subsubsection{\textbf{Clustering Analysis}}
\begin{figure*}[htbp]
\centering
\subfigure[  ]{
\begin{minipage}[t]{0.23\linewidth}
% \vspace{-.1cm}
% \centering
\includegraphics[width=4.5cm]{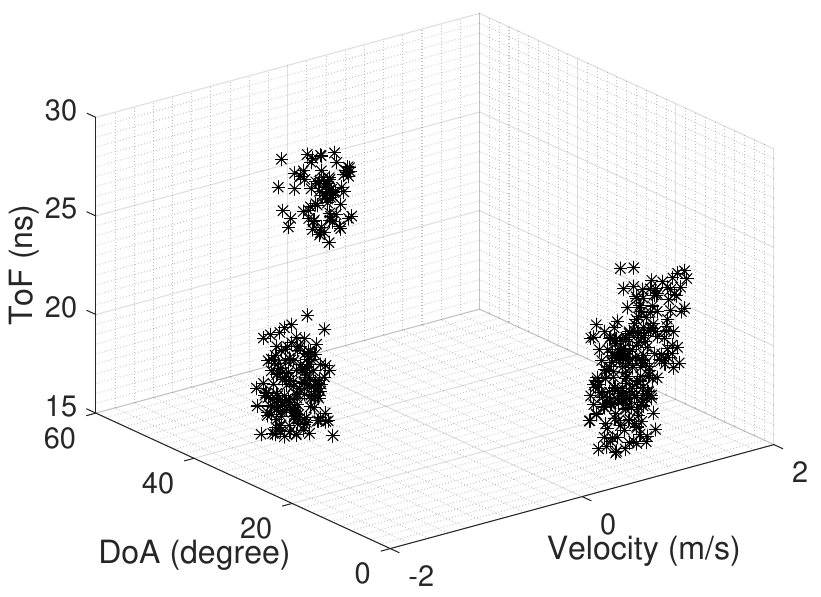}
%\caption{fig2}
\end{minipage}
}%
\subfigure[ ]{
\begin{minipage}[t]{0.235\linewidth}
\centering
\includegraphics[width=4.5cm]{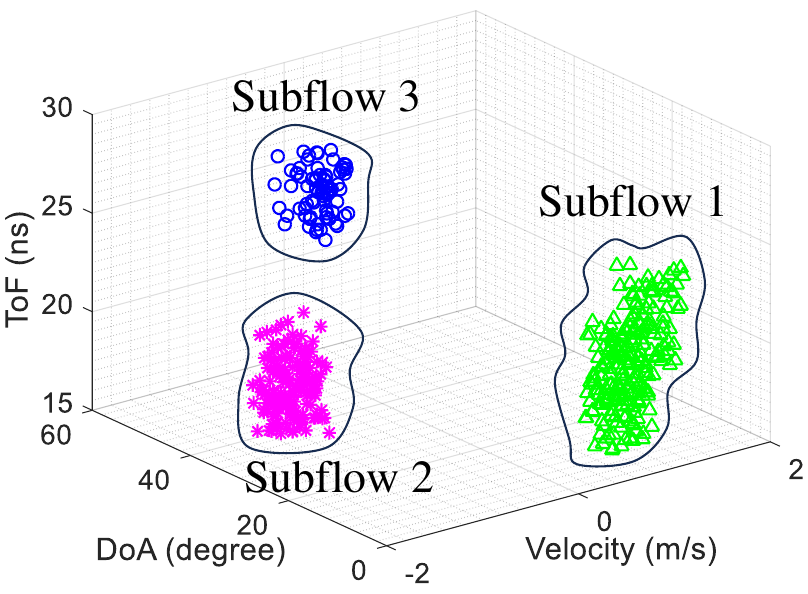}
%\caption{fig2}
\end{minipage}
}%
\subfigure[ ]{
\begin{minipage}[t]{0.23\linewidth}
\centering
\includegraphics[width=4.4cm]{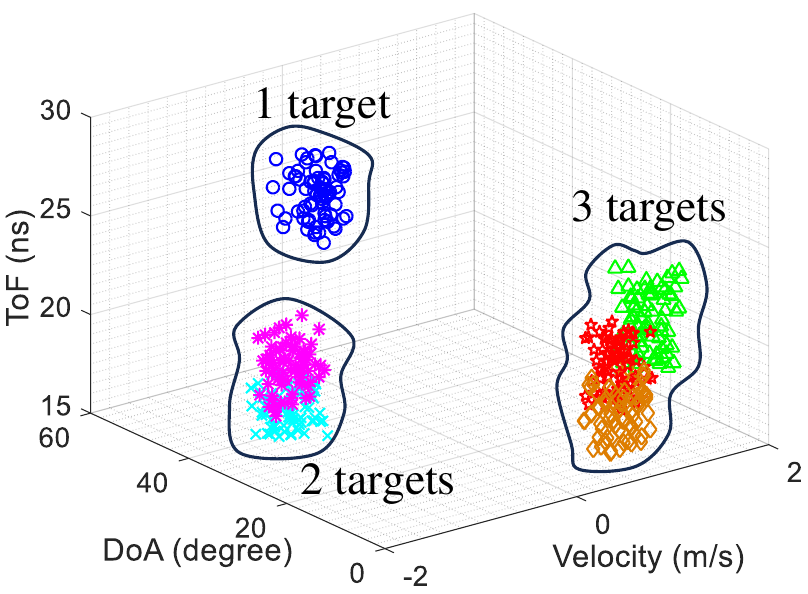}
%\caption{fig2}
\end{minipage}
}
\subfigure[ ]{
\begin{minipage}[t]{0.23\linewidth}
\centering
\includegraphics[width=3.8cm]{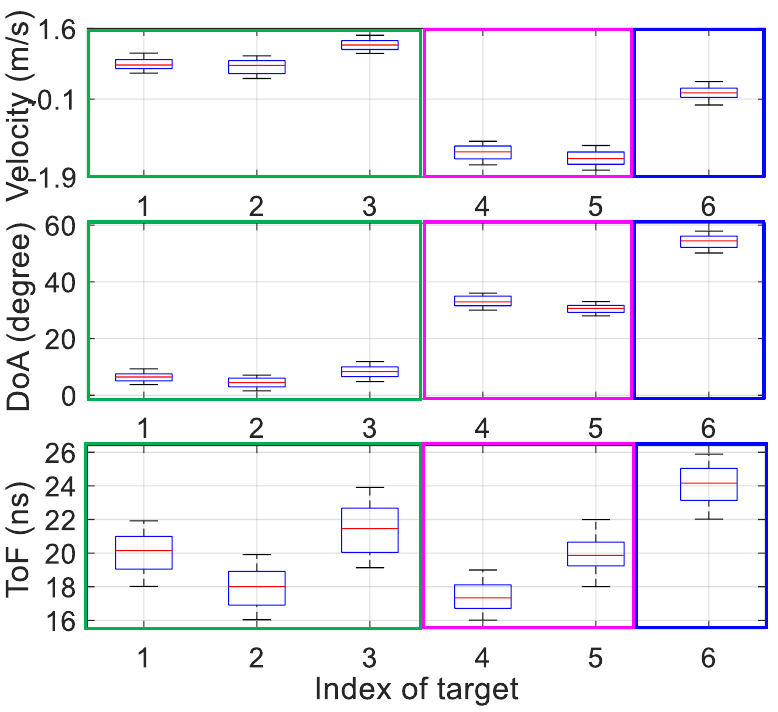}
%\caption{fig2}
\end{minipage}
}
\caption{{The clustering analysis. Figure (a) is the input of clustering, (b) and (c) are the results of affinity propagation clustering and $K$ means based clustering, respectively, and (d) is the parameter distribution.}}
\label{CLST}
\vspace{-0.5cm}
\end{figure*}

Using the obtained signal parameters, we perform the clustering analysis. For instance, with six targets forming three subflows with the sizes of 3, 2, and 1, Fig.~\ref{CLST} shows the clustering outcomes and each target's signal parameter distribution. As can be seen, through the first stage of clustering, G-HFD can first divide the signal parameters into three clusters to get the number of subflows. Then, through the second stage of clustering, G-HFD divides the parameters into 6 clusters. Combining these clustering results, the size of each subflow can be identified. Additionally, as shown in Fig.~\ref{CLST}(d), different subflows display significant differences in DoA and velocity, but less in ToF. This is reasonable, as subflows can move toward different directions at various locations, resulting in more distinct DoAs and velocities. However, their distances relative to the Rx and Tx may be similar, resulting in smaller differences in ToF.

\subsubsection{\textbf{Detecting the Number of Human Targets}}
Based on the denoised V-A spectrum, we investigate the system's performance in detecting the number of targets. During the evaluation, we compare the proposed system with those in~\cite{hao2023toward},~\cite{zhou2020wiflowcount}, and~\cite{liu2022sensor} (denoted as WSTM, WiFlowCount, and SFCC, respectively) and present the results through DA and confusion matrices. \textcolor{black}{Here, the reason for selecting them as comparison systems is that they are all CSI based systems. Moreover, a common feature among them is their utilization of traditional AI techniques for target detection in a typical manner, i.e., signal feature extraction and classification. Specifically, WiFlowCount and SFCC utilize CNN and SVM as classifiers, respectively, to realize detection by classifying extracted signal features. WSTM employs the RNN to capture the spatial-temporal correlations among crowd features for detection. By comparing our system with them, the differences in usage and performance between GAI and traditional AI models in detecting the number of targets can be highlighted.}

\begin{figure}
	\centering
	\subfigure[The detection performance comparison in the corridor.]
    { \begin{minipage}{8cm}
    \includegraphics[width=\textwidth]{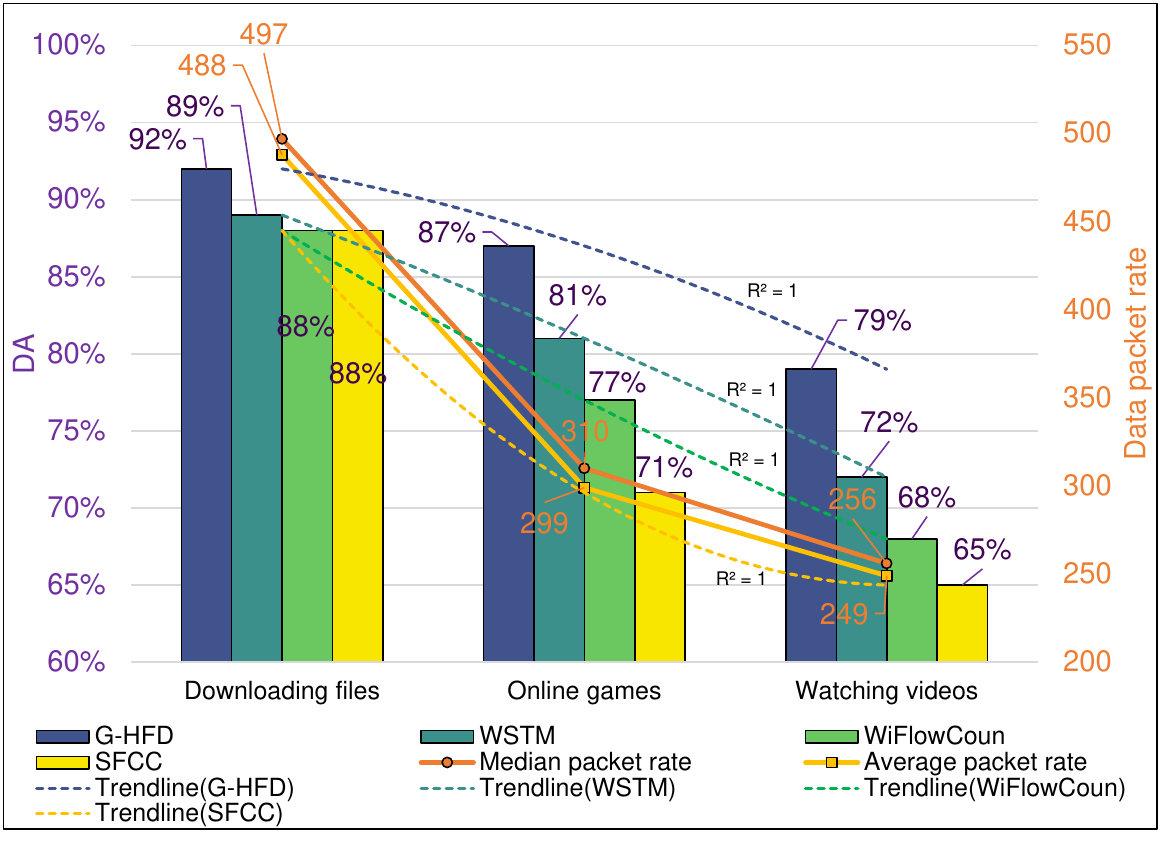}
		\end{minipage}
	}
 \vspace{0.5cm}
	\subfigure[The detection performance comparison in the meeting room.] {
		\begin{minipage}{8cm}
		\includegraphics[width=\textwidth]{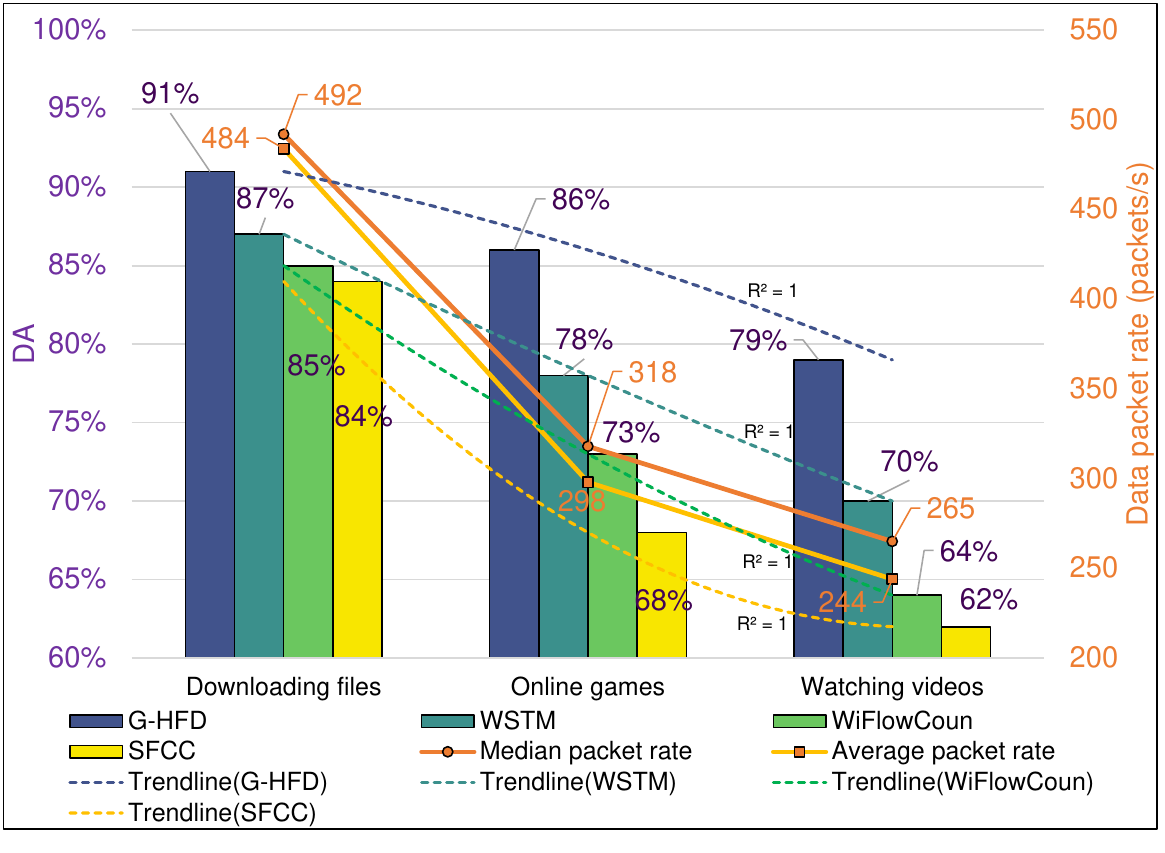}
		\end{minipage}
	}
 \vspace{-0.5cm}
\caption{The performance comparison of detecting the number of targets under two test scenarios.}
\label{NoT-TS}
\end{figure}

\textcolor{black}{Figure~\ref{NoT-TS} presents the performance comparison of detecting the number of targets in two scenarios. For instance, as shown in Fig.~\ref{NoT-TS}(a), when the UE is engaged in downloading files, online games, and watching videos, the DA of the proposed G-HFD is 92\%, 87\%, and 79\%, respectively, which is better than WSTM's 89\%, 81\%, and 72\%, WiFlowCount's 88\%, 77\%, and 68\%, and SFCC's 88\%, 71\%, and 65\%. From the results, we can see that first the G-HFD performs best when UE downloads files, with online games and watching videos following in order. This is because the packet transmission rate of the Tx reaches the highest when UE downloads files, averaging 488 packets per second. This transmission rate decreases to 299 packets per second during online games and further to 249 during watching video. Given the fixed duration covered by $W$ in equation (\ref{eq7}), 0.2 seconds in this paper\footnote{\textcolor{black}{According to the existing research~\cite{wang2023acceleration,goldsmith2005wireless}, in indoor environments, the channel coherence time is approximately 0.2 seconds. Within this time span, the channel parameters can be considered stable.}}, higher transmission rate results in more data for the velocity and acceleration estimation, which boosts the V-A spectrum's resolution and finally enhances the detection accuracy. Similar observations can be made in Fig.~\ref{NoT-TS}(b). For example, when the UE downloads files, the DA of G-HFD can reach 91\%, better than 86\% and 79\%, achieved during online games and watching videos, respectively. Second, for all activities, G-HFD offers better DA than those of the other systems. For example, in a meeting room, the DA of G-HFD can reach 91\% when the UE downloads files, which is better than WSTM's 87\%, WiFlowCount's 85\%, and SFCC's 84\%. The reason is that each moving human target has unique velocity and acceleration. Using the denoised V-A spectrum, G-HFD can distinguish human targets based on the velocity and acceleration, offering a better differentiation than traditional spatial-temporal matrices or Doppler based methods.}

\begin{figure*}[htbp]
\hspace{-0.75cm}
\subfigure[ ]{
\begin{minipage}[t]{0.33\linewidth}
\centering
\includegraphics[width=6.4cm]{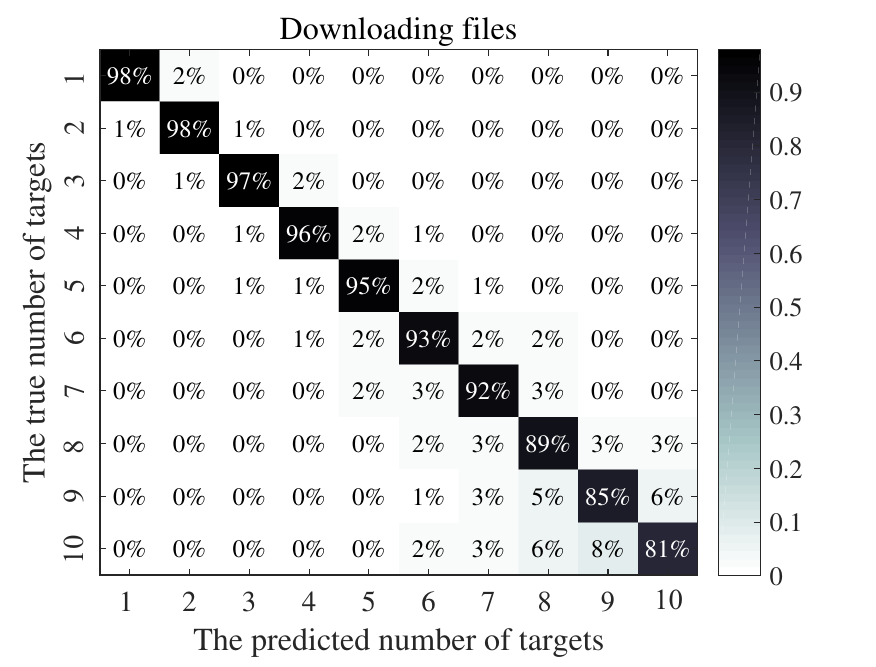}
%\caption{fig2}
\end{minipage}%
}%
\subfigure[ ]{
\begin{minipage}[t]{0.32\linewidth}
\centering
\includegraphics[width=6.4cm]{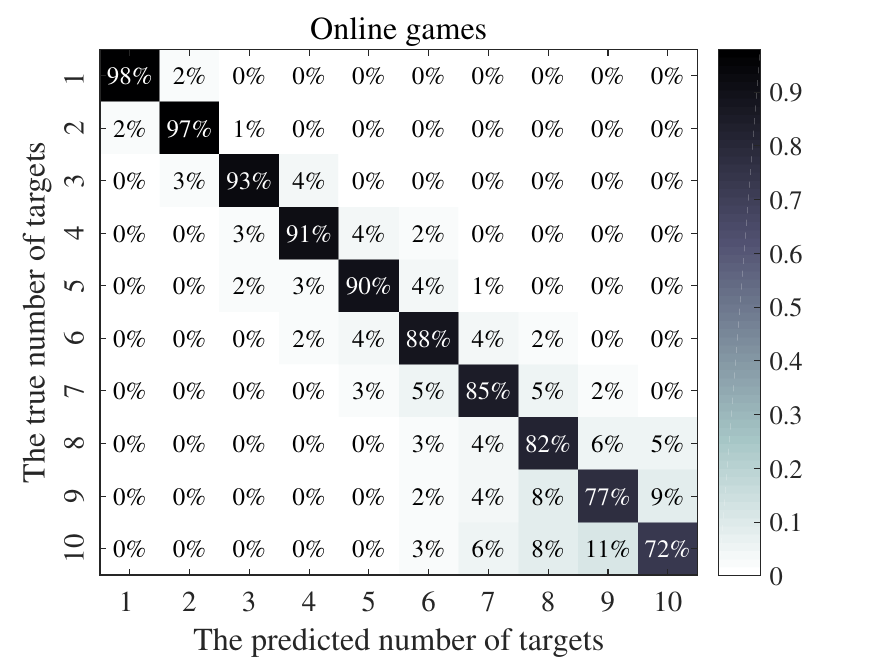}
%\caption{fig2}
\end{minipage}
}%
\subfigure[ ]{
\begin{minipage}[t]{0.3\linewidth}
\centering
\includegraphics[width=6.4cm]{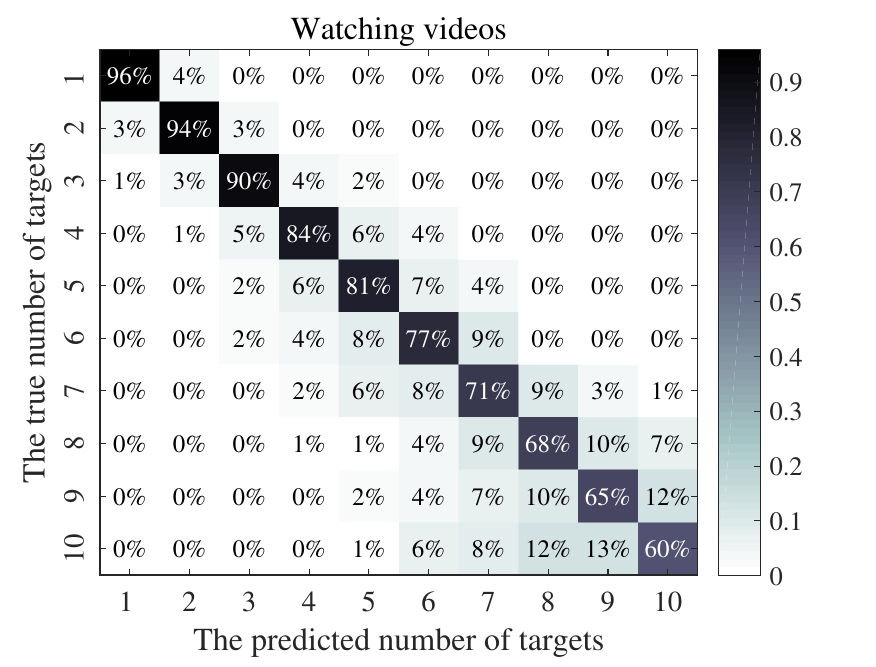}
%\caption{fig2}
\end{minipage}
}%
\centering
\vspace{-0.3cm}
\caption{\textcolor{black}{The confusion matrix of detecting the number of targets in the corridor when UE performs three Internet activities.}}
\label{CFM-C}
\vspace{-0.5cm}
\end{figure*}

\begin{figure*}[htbp]
\hspace{-0.75cm}
\subfigure[ ]{
\begin{minipage}[t]{0.33\linewidth}
\centering
\includegraphics[width=6.4cm]{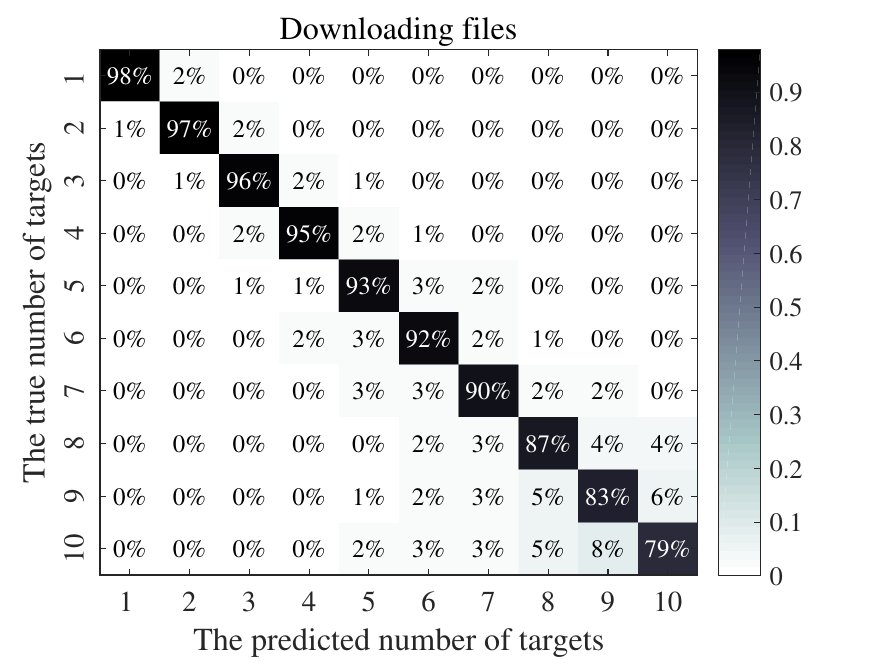}
%\caption{fig2}
\end{minipage}%
}%
\subfigure[ ]{
\begin{minipage}[t]{0.32\linewidth}
\centering
\includegraphics[width=6.4cm]{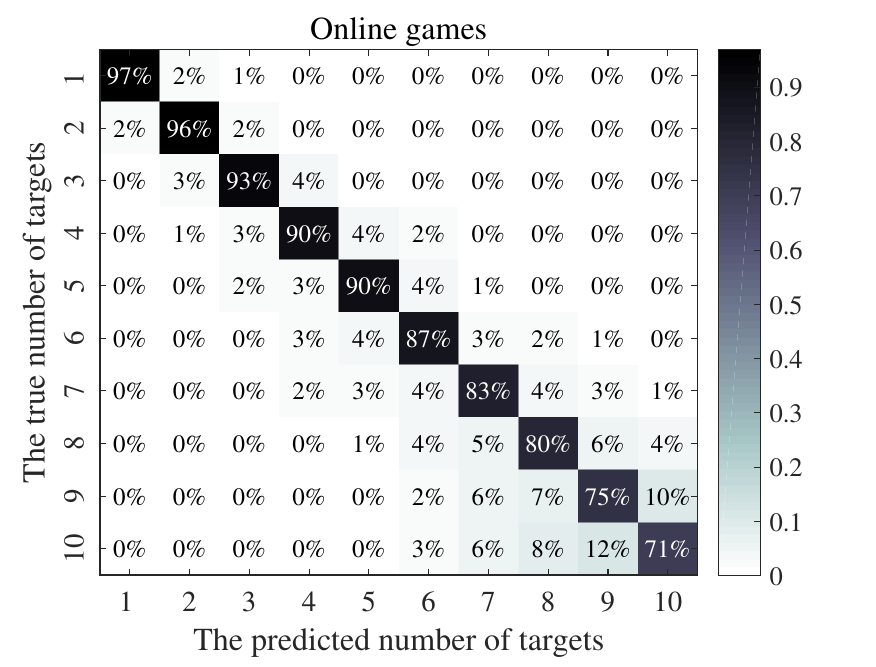}
%\caption{fig2}
\end{minipage}
}%
\subfigure[ ]{
\begin{minipage}[t]{0.3\linewidth}
\centering
\includegraphics[width=6.4cm]{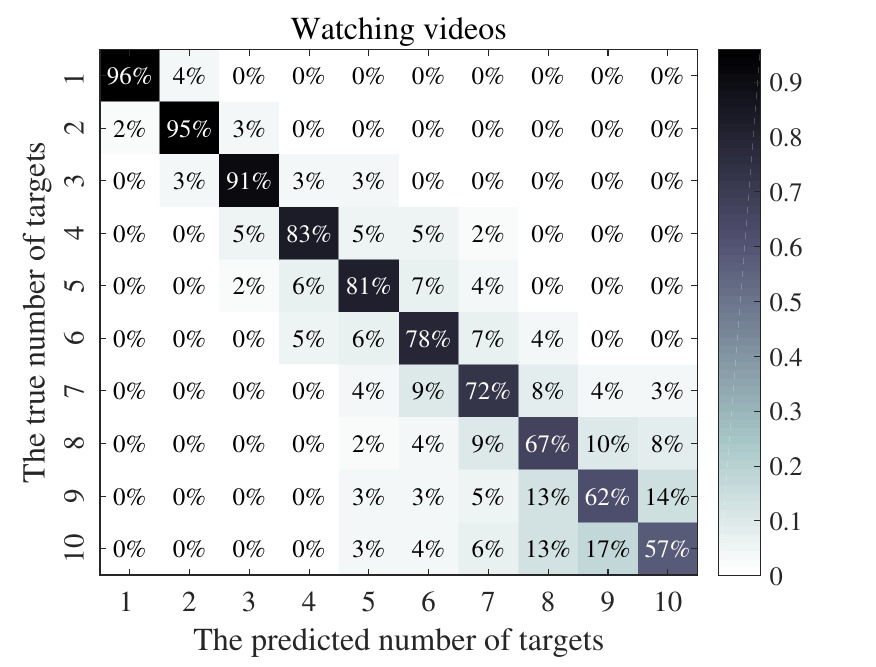}
%\caption{fig2}
\end{minipage}
}%
\vspace{-0.3cm}
\centering
\caption{\textcolor{black}{The confusion matrix of detecting the number of targets in the meeting room when UE performs three Internet activities.}}
\label{CFM-M}
\end{figure*}

\textcolor{black}{From the confusion matrix in Fig.~\ref{CFM-C} and Fig.~\ref{CFM-M}, we can see that in both scenarios, the DA decreases with an increase in the number of human targets. Specifically, in the corridor, taking the downloading files as an example, the DA of G-HFD decreases from 98\% to 81\% as the number of targets increases from 1 to 10. This deterioration is more severe for the Internet activities with a lower packet transmission rate, such as watching videos. For instance, in the meeting room, when the UE is engaged in watching videos, G-HFD's DA decreases from 96\% to 57\% as the number of targets increases from 1 to 10. The decline is attributed to the increased mutual interference among more dynamic targets, which introduces additional noise into the V-A spectrum, adversely affecting detection performance. Furthermore, the system's performance is more significantly affected when the packet transmission rate is lower. More fundamentally, G-HFD's detection of human targets relies on the V-A spectrum, which means the maximum number of targets that can be detected is bounded by both the resolution and the noise level of the V-A spectrum. Here, the resolution is determined by the packet transmission rate, while the noise level is influenced by various factors, including the physical environment and the presence of human targets. For a specific number of targets, a lower transmission rate results in lower V-A spectrum resolution and DA. Similarly, with fixed transmission rate (i.e., V-A resolution), the detection performance is mainly affected by the noise level. Therefore, increasing the packet transmission rate appropriately can enhance the detection capacity for the maximum number of targets.}

% \vspace{-1cm}

\subsubsection{\textbf{Subflow Detection }}
\begin{figure*}[htbp]
\hspace{-0.75cm}
\subfigure[ ]{
\begin{minipage}[t]{0.33\linewidth}
\centering
\includegraphics[width=6cm]{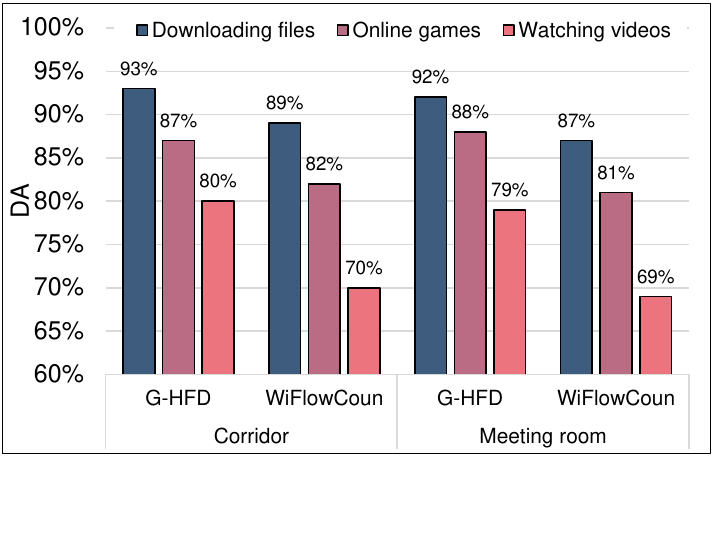}
%\caption{fig2}
\end{minipage}%
}%
\subfigure[ ]{
\begin{minipage}[t]{0.32\linewidth}
\centering
\includegraphics[width=6.2cm]{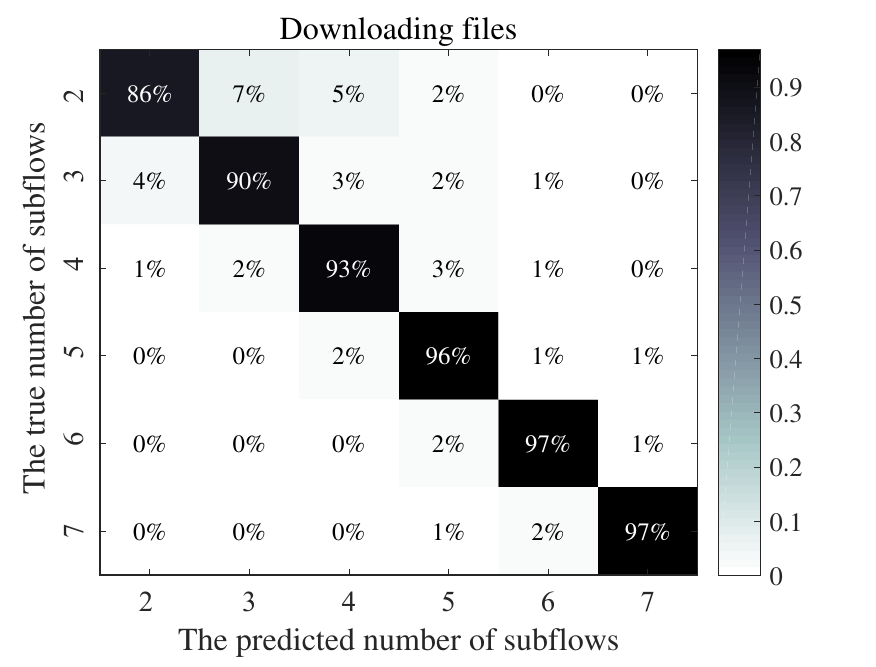}
%\caption{fig2}
\end{minipage}
}%
\subfigure[ ]{
\begin{minipage}[t]{0.3\linewidth}
\centering
\includegraphics[width=6.2cm]{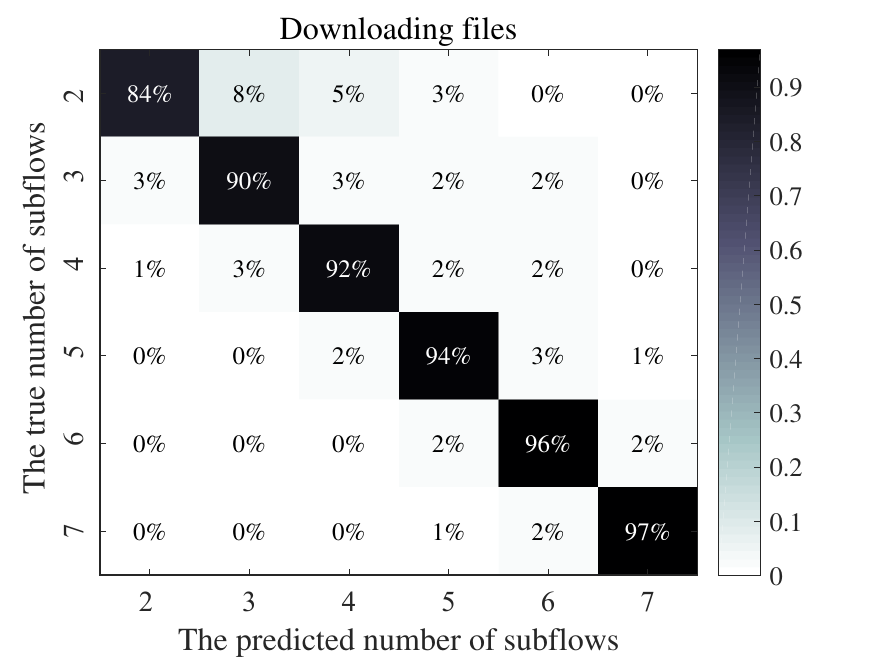}
%\caption{fig2}
\end{minipage}
}%
\vspace{-0.3cm}
\centering
\caption{\textcolor{black}{The performance comparison of detecting the number of subflows. (a) shows the comparison of overall detection accuracy. (b) and (c) display the confusion matrices for detecting the number of subflows in the corridor and meeting room, respectively, when the UE downloads files.}}
\label{SUBF}
\end{figure*}

Finally, we evaluate the G-HFD's performance in detecting the number of subflows and the size of each subflow in two test scenarios involving ten dynamic targets. \textcolor{black}{Figure~\ref{SUBF} illustrates the performance of detecting the number of subflows. For instance, in the corridor, the DAs of G-HFD are 93\%, 87\%, and 80\%, respectively, outperforming WiFlowCount's 89\%, 82\%, and 70\%, when UE performs three different activities. This implies that G-HFD can effectively separate target induced signals through clustering in the space formed by velocity, DoA, and ToF, thereby accurately identifying the number of subflows. In comparison to WiFlowCount based on the Doppler effect, G-HFD characterizes the signal parameters of HIR from multiple dimensions, which is more comprehensive, thereby performing better. On this basis, the confusion matrices reveal that G-HFD's DA drops when there are fewer subflows. Concretely, as can be seen in Figs.~\ref{SUBF}(b) and (c), when the number of subflows decreases from 7 to 2, the DA declines from 97\% to 86\% in the corridor. Similarly, in the meeting room, the DA decreases from 97\% to 84\%, further demonstrating the effect of reducing the number of subflows on the detection performance. This can be interpreted by that, with fixed number of targets, fewer subflows mean more targets in each subflow, resulting in more data points corresponding to each subflow. These points cover wider range across three dimensions, making them likely to be wrongly clustered into multiple groups, which finally reduces the DA.}

\begin{figure*}[htbp]
\hspace{-0.75cm}
\subfigure[ ]{
\begin{minipage}[t]{0.33\linewidth}
\centering
\includegraphics[width=6cm]{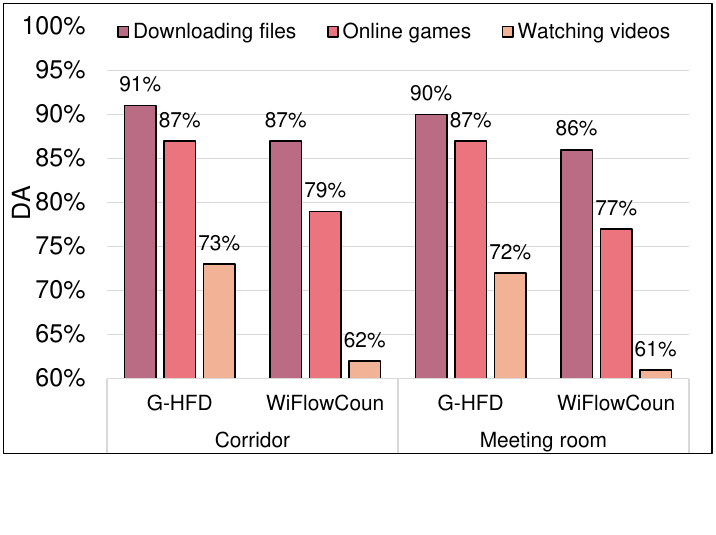}
%\caption{fig2}
\end{minipage}%
}%
\subfigure[ ]{
\begin{minipage}[t]{0.32\linewidth}
\centering
\includegraphics[width=6.2cm]{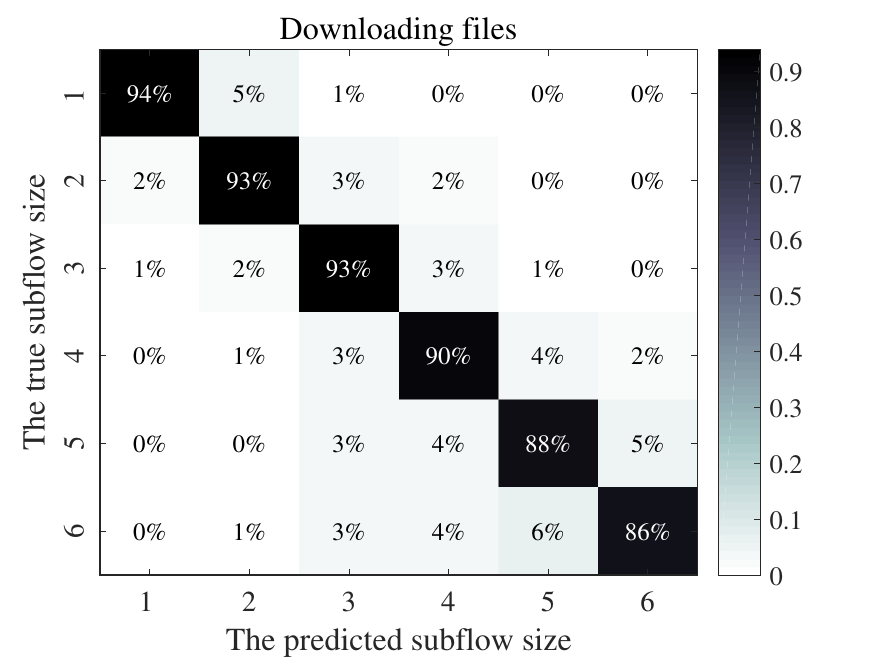}
%\caption{fig2}
\end{minipage}
}%
\subfigure[ ]{
\begin{minipage}[t]{0.3\linewidth}
\centering
\includegraphics[width=6.2cm]{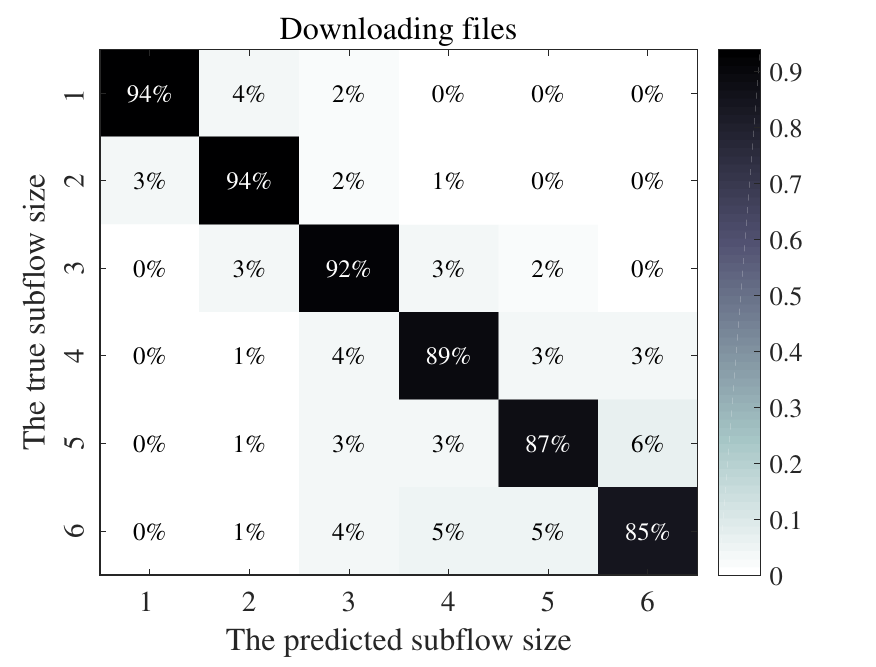}
%\caption{fig2}
\end{minipage}
}%
\vspace{-0.3cm}
\centering
\caption{\textcolor{black}{The performance comparison of subflow size detection. (a) shows the comparison of overall detection accuracy. (b) and (c) display the confusion matrices for detecting the subflow size in the corridor and meeting room, respectively, when the UE downloads files.}}
\label{SUFS}
\end{figure*}

Figure~\ref{SUFS} presents the subflow size detection performance of G-HFD. \textcolor{black}{As can be seen, for the three different UE activities, the DAs of G-HFD in the corridor are 91\%, 87\%, and 73\%, respectively, better than WiFlowCount's 87\%, 79\%, and 62\%. In the meeting room, G-HFD's DA can reach 90\%, 87\%, and 72\%, while WiFlowCount can reach 86\%, 77\%, and 61\%, respectively.} These results demonstrate the effectiveness of G-HFD in subflow size detection. Meanwhile, they also reveal that G-HFD provides higher DA. This superiority fundamentally stems from G-HFD's reliance on analyzing features of the target induced reflections across time, space, and frequency domains for detection. Compared with the Doppler spectrum, these features depict the spatial distribution and dynamic characteristics of the target more accurately and comprehensively, thereby enhancing its performance.

\textcolor{black}{Additionally, the confusion matrices show a decrease in DA as subflow size increases. For instance, in the corridor, the DA decreases from 94\% to 86\% when the subflow size increases from 1 to 6. Similarly, for the meeting room, we can see that the DA drops from 94\% to 85\%. This occurs because fewer subflows with fixed number of targets means more targets per subflow, causing two negative impacts. First, the mutual influence among dynamic targets within the same subflow becomes more significant, which amplifies the noise in the V-A spectrum, affecting the detection of the number of the targets. Second, the range of signal parameters covered by each subflow expands, resulting in potential overlaps of signal parameters from different subflows, which impacts the recognition of the number of subflows. Consequently, this degradation in detecting the number of targets and subflows adversely affects subflow size detection performance. For such performance degradation, the fundamental solution lies in enhancing the differentiation between data points corresponding to different targets, which depends on the signal parameter estimation accuracy. Hence, we can mitigate such degradation by increasing the signal bandwidth, packet transmission rate, and adding more antennas to improve parameter estimation performance.}

\section{Conclusion}
This paper proposes G-HFD, which, unlike current traditional AI model-based systems, leverages GAI to extract and enhance the signal parameters, enabling fine-grained flow detection, including the total number of human targets and subflows, as well as subflow sizes. Within G-HFD, we propose the diffusion model-based UW-CDM, which can denoise the velocity and acceleration estimation results, thereby improve the system's ability to detect the number of human targets. Additionally, the UW-CDM can also generate the clear DoA spectrum based on a given ambiguous one, which enables the DoA estimation when the antenna spacing exceeds half a wavelength and facilitates the subflow detection. Using the downlink signals in practical communication, the evaluations show that G-HFD's human subflow size detection accuracy can reach 91\%, when the UE downloads files. This not only validates G-HFD's effectiveness in fine-grained flow detection, but also demonstrates the crucial potential of GAI in wireless sensing. In the future, we will further explore the applications of GAI models in signal processing and their potential support for wireless sensing.
% \vspace{-.3cm}
\bibliographystyle{IEEEtran}
% \bibliography{references.bib}
\bibliography{Ref.bib}

% \end{IEEEbiography}
\end{document}